\documentclass{aa}
\usepackage{graphicx}
\usepackage{txfonts}
\usepackage[normalem]{ulem}
\usepackage{hyperref}
\hypersetup{
  colorlinks,
  citecolor=blue,
  linkcolor=blue,
  urlcolor=blue}

\defcitealias{Ertl2016}{E16}
\defcitealias{M16}{M16}
\defcitealias{Schneider21}{S21}
\defcitealias{Schneider2023}{S23}
\defcitealias{Schneider2024}{S24}
\defcitealias{Temaj}{T24}
\defcitealias{Fryer2012}{F12}
\defcitealias{Mandel2020}{MM20}
\defcitealias{Mapelli2020}{M20}
\defcitealias{PS20}{PS20}

\begin{document}

    \title{Explodability criteria for the neutrino-driven 
    supernova mechanism}
    \titlerunning{Explodability criteria for the neutrino--driven
    supernova mechanism}

    \author{
        K.~Maltsev\inst{1,2}\and
        F.~R.~N.~Schneider\inst{1,3}\and
        I.~Mandel\inst{4,5} \and
        B.~Müller\inst{4} \and
        A.~Heger\inst{4} \and
        F.~K.~R{\"o}pke\inst{1,2} \and
        E.~Laplace\inst{1,6,7}
     }
    \institute{
        Heidelberger Institut f{\"u}r Theoretische Studien,
        Schloss-Wolfsbrunnenweg 35, D-69118 Heidelberg, Germany\\
        \email{kiril.maltsev@protonmail.com}
        \and
        Zentrum f\"ur Astronomie der Universit\"at Heidelberg, Institut f\"ur Theoretische Astrophysik, Philosophenweg 12, D-69120 Heidelberg, Germany
        \and
        Zentrum f\"{u}r Astronomie der Universit\"{a}t Heidelberg, Astronomisches Rechen-Institut, M\"{o}nchhofstr. 12-14, D-69120 Heidelberg, Germany
        \and
        School of Physics and Astronomy, Monash University, Clayton, Australia
        \and
        OzGrav: The ARC Center of Excellence for Gravitational Wave Discovery, Australia
        \and
        Institute of Astronomy, KU Leuven, Celestijnenlaan 200D, B-3001 Leuven, Belgium
        \and
        Anton Pannekoek Institute of Astronomy, University of Amsterdam, Science Park 904, 1098 XH Amsterdam, The Netherlands }

    \date{Received 31 March 2025 / Accepted 20 May 2025}
    
    \abstract 
    { 
      Massive stars undergoing iron core-collapse at the end of their evolution terminate their lives either as successful or failed supernovae (SNe). The physics of core-collapse supernovae (CCSNe) is complex and understanding it requires computationally expensive simulations. Therefore, using these simulations to predict CCSN outcomes over large, densely sampled parameter spaces of SN progenitors (as is needed, e.g., for population synthesis studies) is not feasible. To remedy this situation, we present a set of explodability criteria that allow us to predict the final fates of stars by evaluating multiple stellar structure variables at the onset of core collapse. These criteria are calibrated to predictions made using a semi-analytical SN model, when evaluated over a set of $\sim$~3,900 heterogeneous stellar progenitors (comprised of single stars, binary-stripped and accretor stars).  Over these, the explodability criteria achieve a >99\% agreement with the semi-analytical model. We test these criteria  on 
      29 state-of-the-art 3D CCSN simulation outcomes from two different groups. 
      
      Furthermore, we find that all explodability proxies needed for our pre-SN structure-based criteria
      have two distinct peaks and intervening valleys as a function of the carbon-oxygen (CO) core mass, $M_\mathrm{CO}$, coinciding with failed and successful SNe, respectively. The CO core masses of explodability peaks shift systematically with metallicity, $Z$, as well as with the timing of the hydrogen-rich envelope removal via binary mass transfer. Using the criteria and the systematic shifts, we identify critical values in $M_\mathrm{CO}$ that define windows over which black holes form by direct collapse. With these findings, we formulate a CCSN recipe based on $M_\mathrm{CO}$ and $Z$ that is applicable for rapid binary population synthesis and other studies. Our explodability formalism is consistent with observations of Type~IIP, IIb, Ib and Ic SN progenitors and it partially addresses the missing red supergiant problem by direct black hole formation. 
    }
    

    \maketitle

\section{Introduction}\label{sec:introduction}
When the iron core of a massive star surpasses its effective Chandrasekhar mass \citep{Timmes1996}, it collapses. During collapse, the core density grows until it reaches nuclear densities and a proto-neutron star (PNS) forms. The still infalling outer core bounces off the PNS, launching an outwardly propagating shock wave. The shock wave loses momentum due to the photodissociation of the heavy nuclei and stalls \citep{Colgate1966, Bethe1985}. The leading paradigm\footnote{In this work, we consider the neutrino-driven engine and not alternative SN mechanisms, such as those driven by magnetars \citep{Kasen2010}, collapsars \citep{Woosley2006}, or jittering jets \citep{Papish2011}.} is that neutrino heating imparts the energy that facilitates shock revival in the case of a successful SN explosion. Since asymmetry-induced turbulence in both the progenitor structure at the onset and during the collapse, as well as other multidimensional effects, can be critical to shock revival \citep[e.g.,][]{Mueller2020}, the appropriate modeling of core-collapse supernovae (CCSNe) ought to be addressed using three-dimensional (3D) simulations. However, 3D CCSN simulations are notoriously expensive computationally due to the spatial and temporal scales that need to be resolved, along with neutrino transport and the complexity of other physical processes involved \citep{JANKA_2007, Burrows, 2025janka}.
For population synthesis purposes that demand modeling hundreds or thousands of stars, only one-dimensional (1D) codes of stellar evolution up to core collapse and only 1D CCSN codes are computationally feasible. 

To overcome these challenges, different explodability proxies, derived from the pre-SN stellar structure, have been introduced, based on large samples of 1D CCSN simulation outcomes. These are used 
to predict whether iron core collapse will result in success or failure of the neutrino-driven SN mechanism. 
The question of whether stars can explode by the neutrino energy transport mechanism is linked to the
density structure outside the iron core \citep{Burrows1987}.  \cite{OConnor2021} introduced the compactness parameter 
\begin{equation}
    \xi_M = \frac{M/M_\odot}{R(M)/1000 \, \mathrm{km}}
\end{equation}
as a first-order criterion for predicting the final fates. Here, $R$ is the radius of the Lagrangian mass shell enclosing a baryonic mass, $M$, in the pre-SN star. 
For $M \, {=} \, 2.5 \, M_\odot$, they found that setting the condition $\xi_{2.5} \, {>} \, 0.45$ for a failed SN is in agreement with CCSN outcomes predicted with the GR1D SN code \citep{OConnor2010} over a sample of more than 700 simulations for 100 different progenitor models. 
Based on 2D and 3D CCSN simulations with approximate neutrino transport, \cite{Horiuchi2014} concluded that a lower threshold of $\xi_\mathrm{2.5} > 0.2$ better represents the CCSN outcomes in their sample and is in line with observational constraints. Many other works confirmed the following gross trend, based on different sets of stellar progenitor models and various SN codes: stars with lower $\xi_\mathrm{2.5}$ are statistically more likely to explode \citep[e.g.,][]{OConnor2013, Nakamura2015, Sukhbold2018, Schneider21, Takahashi,Laplace2025}. 

However, with the \textsc{Prometheus-hotbath} 1D hydrodynamics code that parametrizes the contraction dynamics of the PNS, 
\cite{Ugliano2012} obtained successful SNe for $\xi_\mathrm{2.5}<0.15$, failed SNe for $\xi_\mathrm{2.5}>0.35$, and their coexistence for values in between. \cite{Pejcha2015} found a similar degeneracy in the final fate outcomes over a large set of SN progenitors, even when optimizing the choice of $M$ for $\xi_M$, while comparing the neutrino luminosity obtained from GR1D simulations to the analytically estimated critical neutrino luminosity required for shock revival. It has further  been  criticized that a high compactness does not consistently coincide with failed SNe in multidimensional CCSN simulations and, instead, it often anticipates the most energetic SN explosions \citep{burrows2024, Wang2022, Burrows2019}.

As an alternative, \cite{Ertl2016} made the choice of the mass coordinate dependent on an entropy condition, and introduced a two-parameter criterion for classifying explodability, based on properties of the normalized mass,
\begin{equation}
    M_4 = \frac{m (s=4 \, k_B)}{M_\odot},
\end{equation}
which is located where the entropy per baryon is $s \, {=} \, 4 \, k_B$. At this mass coordinate, the entropy abruptly rises and the density declines. It typically defines the mass shell of the PNS in whose vicinity the shock is revived in the case of a successful SN explosion. 
Smaller values of the radial gradient at the mass coordinate $M_4$, 
\begin{equation}
    \mu_4 =  \left.\frac{\mathrm{d}m/M_\odot}{\mathrm{d}r/1000 \mathrm{km}} \right \vert_\mathrm{s=4},
\end{equation}
imply a steeper density jump; thus,  at core-collapse, lower-density matter arrives at the neutrino-sphere and the ram pressure is then  reduced \citep{Sukhbold2018}. 
The hot accretion mantle pushes onto the PNS, giving rise to a neutrino luminosity that is maintained by persistent mass accretion. This luminosity, $L_{\nu}^\mathrm{acc} \propto G \, M_\mathrm{PNS} \, \dot{M} / R_\mathrm{PNS}$, depends on the mass accretion rate, $\dot{M}$, on the PNS mass $M_\mathrm{PNS} \sim M_4$, and radius, $R_\mathrm{PNS}$. 
For neutrino luminosities above a critical threshold, $L_{\nu, \mathrm{crit}}$, namely,\begin{equation}
\label{eq:nu_cond}
    L_{\nu}^\mathrm{acc} > L_{\nu, \mathrm{crit}},
\end{equation}
the neutrino heating triggers the onset of an explosion by shock runaway expansion \citep{Burrows1993}. 
The variable $\mu_4$, when divided by the free-fall timescale and multiplied by the radius up to which $M_4$ is enclosed, is proportional to $\dot{M}$ \citep{Ertl2016}. Since $R_\mathrm{PNS}$ is similar across different progenitors and only weakly time-dependent in late-time explosions, the product $\mu_4 M_4 $ is a proxy for $L_{\nu}^\mathrm{acc}$ \citep{Ertl2016}. Based on comparison to CCSN outcomes obtained with the \textsc{prometheus-hotbath} code, the authors found that a separation line in the $(\mu_4 M_4, \mu_4)$ plane segregates progenitors that will fulfill the condition~(\ref{eq:nu_cond}) from those that will not. The $(\mu_4 M_4, \mu_4)$ plane compares two competing forces onto the CCSN outcome: failed SNe are favored by a high density outside of the iron core (i.e., a large $\mu_4$) and by weak accretion luminosities (i.e., a small $\mu_4 M_4$ compared to $\mu_4$). Thus, SN progenitors below the separation line are predicted to explode (and to implode otherwise). With this explodability criterion, which we refer to as \citetalias{Ertl2016},
they achieved an accuracy of $\simeq 97 \%$ over a heterogeneous set of 621 massive single star progenitors, as predicted by a 1D CCSN model at a reduced neutrino wind power compared to the approach of \cite{Ugliano2012}. 

While a two-parameter criterion is more appropriate to model explodability \citep[see, e.g.,][]{Heger2023}, a separation line in the $(\mu_4M_4, \mu_4)$ plane has not been shown to be a reliable criterion to discriminate the outcomes of other SN codes such as \textsc{stir} \citep{Couch2020} -- which models the effects of turbulence in 1D by a modified mixing-length theory approach\footnote{However, see \cite{Mueller2019mnras} for a critical assessment of this approach to incorporating the effects of turbulence.} -- or the 2D \textsc{fornax} code \citep{Tsang2022}.

\citetalias{Ertl2016} assesses the shock revival conditions at one instant of time, namely, at the onset of iron-core infall. It therefore does not capture the temporal-dynamical nature of how the heating and accretion conditions compare as the collapse proceeds. One of the original motivations for the formulation of a semi-analytical SN model in \cite{M16}, which we refer to as \citetalias{M16}, was to supply a set of ordinary differential equations that takes the dynamical evolution of these competing effects into account. \citetalias{M16} draws together theoretical insights, scaling relations, and analytical approaches from previous works into a unified framework that models the neutrino-driven perturbation-aided SN engine.  To predict whether the stalled shock is revived or not, \citetalias{M16} estimates the evolution of the mass advection ($\tau_\mathrm{adv}$) and the neutrino heating ($\tau_\mathrm{heat}$) timescales and evaluates whether the critical condition for success of the neutrino-driven SN mechanism \citep{Janka1998, Janka2000},
\begin{equation}
    \tau_\mathrm{adv} > \tau_\mathrm{heat},
\end{equation}  
is eventually fulfilled during core-collapse.
For a discussion of alternative critical conditions, such as the ``antesonic'' \citep{Pejcha2012} and the ``force explosion'' \citep{Murphy2017} conditions that all are expected to result in Eq.~(\ref{eq:nu_cond}), see \cite{Pochik2024}. 
The critical condition for the onset of an explosion during core collapse is distinct from the progenitor problem, which we address in this work. The latter aims to discriminate whether pre-SN models will fulfill a critical condition (and thus explode) or not.

The \citetalias{M16} model outcomes have been compared to those of the $\xi_\mathrm{2.5}$-based and \citetalias{Ertl2016} criteria over the same set of SN progenitors \citep[e.g.,][]{Sukhbold2018, Schneider2023, Takahashi, Aguilera-Dena}; however, no simple \citetalias{M16} SN outcome-progenitor connection based on a reduced set of pre-SN variables has been established thus far.\footnote{
Other explosion conditions based on pre-SN stellar structure profiles have been introduced, such as those in \cite{Boccioli2023} (based on 1D simulations with GR1D) and in \cite{Wang2022} (based on 2D simulations with \textsc{fornax}), which stress the importance of a steep density jump at the interface between the silicon core and the silicon-enriched oxygen layers for a successful explosion. Their analysis and comparison to \citetalias{M16} is beyond the scope of this work.}
To this end, we formulate a set of explodability criteria for the neutrino-driven SN mechanism using multiple diagnostic pre-SN stellar structure variables: $\xi_\mathrm{2.5}$, the \citetalias{Ertl2016} parameters $\mu_4$ and $\mu_4 M_4$, the carbon-oxygen core mass $M_\mathrm{CO}$, and the central specific entropy, $s_c$.  We calibrate  these criteria to CCSN outcomes predicted by \citetalias{M16} 
over a large sample of $\simeq 3900$ single, binary-stripped and accretor star 1D models\footnote{
Detailed 3D simulations provide a more reliable and direct way to relate SN outcomes to progenitor properties than 1D models or \citetalias{M16}, which naturally fall short of capturing the multidimensional nature of CCSNe. However, larger suites of 3D simulations  needed for this purpose are currently not available.
}.

We then bridge the gap towards their application for rapid binary population synthesis (BPS). Since \citetalias{M16} (and 1D CCSN  models) 
require the entire progenitor structure profiles at the pre-SN stage as input, these models cannot be applied for BPS for two reasons. First, 
BPS codes use stellar models that are evolved only up to a cutoff at (if not before) the central neon ignition. Second, rapid BPS codes 
do not keep track of entire stellar structure profiles 
and only evolve a set of global parameters. 
To predict the final fates of massive stars undergoing iron core collapse, rapid codes use recipes such as those introduced in \cite{Fryer2012}, \cite{Mapelli2020}, or \cite{Mandel2020}, which are all based on $M_\mathrm{CO}$. In this work, we use our pre-SN explodability criteria and a set of stellar models to construct a $M_\mathrm{CO}$-based CCSN recipe that distinguishes single and binary-stripped stars and takes the metallicity ($\mathrm{Z}$) dependence of explodability into account.

This paper is structured as follows. In Sect.~\ref{sec:meth}, we introduce our CCSN models and the stellar SN progenitors. The pre-SN explodability criteria telling apart successful and failed SNe are formulated and extended by a scheme predicting the remnant type 
in Sect.~\ref{sec:shock-revival-univ-preSN}. The compact remnant type landscape of single stars and binary-stripped stars, predicted using our CCSN recipe, is the main result presented in Sect.~\ref{sec:ff_land}. 
We compare our pre-SN explodability criteria and the $M_\mathrm{CO}$-based CCSN recipe with 3D CCSN simulation outcomes and with other SN models in Sect.~\ref{sec:comp_3d}. In Sect.~\ref{sec:comp_obs}, our and alternative CCSN recipes are compared against observations that constrain the SN progenitor properties. The results are discussed in Sect.~\ref{sec:disc} and our conclusions are drawn in Sect.~\ref{sect:con}.

\section{Methods}
\label{sec:meth}

In Sect.~\ref{sec:prog-m}, we introduce the set 
of single, binary-stripped and accretor star SN progenitor models, 
the \citetalias{M16} CCSN model in Sect.~\ref{sec:sn-m}, and the catalog of 3D CCSN simulation used in this work in Sect.~\ref{sec:3d-models}. In Sect.~\ref{sec:sl-m}, we refer to
the supervised learning model that
we use to map out CCSN outcomes as a function of progenitor $M_\mathrm{CO}$. 

\subsection{1D CCSN progenitor models}
\label{sec:prog-m}

We compile the following set of stellar evolution models:
\begin{itemize}
    \item 127 single stars and binary-stripped stars [Case A, Case Be, Case Bl, Case C] of variable zero-age main sequence (ZAMS) mass $11 \leq M_\mathrm{ZAMS}/M_\odot \leq 90$ at a solar\footnote{As solar metallicity, we assume $Z_\odot = 0.001432,$ following \cite{Asplund2009}.} metallicity of $Z = Z_\odot$ from \cite{Schneider21}, which we refer to as \citetalias{Schneider21},
    \item 121 single stars and binary-stripped stars [Case A, Case Be, Case Bl, Case C] of variable $11 \leq M_\mathrm{ZAMS}/M_\odot \leq 90$ at $Z = Z_\odot/10$ from \cite{Schneider2023}, which we refer to as \citetalias{Schneider2023},
    \item 570 accretor stars [Case Ae\footnote{The distinction between Case~Ae and Case~Al is made depending on whether the mass transfer occurs before (Ae) or after (Al) the mid--MS.}, Case~Al, Case~Be, Case~Bl, Case~C] of variable $11 \leq M_\mathrm{ZAMS}/M_\odot \leq 90$ and fraction $f \in (0.1, 2)$ of ZAMS mass accreted on the thermal timescale at $Z = Z_\odot$ from \cite{Schneider2024}, which we refer to as \citetalias{Schneider2024}, 
    \item 169 single stars of variable $9 \leq M_\mathrm{ZAMS}/M_\odot \leq 70$ and convective core overshooting parameter $0.05 \leq \alpha_\mathrm{ov}/H_P \leq 0.5$ at $Z = Z_\odot$ from \cite{Temaj}, which we refer to as \citetalias{Temaj} and 
    \item 2910 single stars of variable $9 \leq M_\mathrm{ZAMS}/M_\odot \leq 45$ at $Z = Z_\odot$ from \cite{M16}, which we refer to as H16.
\end{itemize}
All 3897 stellar models were evolved from ZAMS up to the onset of iron-core infall. 
The first four data sets from \citetalias{Schneider21}, \citetalias{Schneider2023}, \citetalias{Schneider2024}, and \citetalias{Temaj} have in common the following: for the advanced burning phases, the same input physics is assumed; and these were evolved using the \textsc{mesa} code \citep{Paxton2011, Paxton2013, paxton2015, Paxton2018, Paxton2019}. H16 adopted a different input physics for the main and the advanced burning phases and the stellar models were evolved using the \textsc{Kepler} code \citep{Weaver1978, Heger2010}.

The classification of binary-stripped stars in \citetalias{Schneider21} and \citetalias{Schneider2023} is based on the mass transfer (MT) history, following the nomenclature summarized in Table~\ref{Tab:cases}. The stripped stars are modeled effectively as single stars, with a prescription for removal of the hydrogen-rich envelope over 10\% of the thermal timescale. For details on the effective modeling approach, we refer to \cite{Schneider21}.

\begin{table}
  \begin{center}
  \caption{\label{Tab:cases}Classification of binary-stripped and accretor star models based on the mass transfer history.}
    \begin{tabular}{l l}       
    \hline\hline 
       MT class & Timing of hydrogen-rich envelope removal \\
      \hline
      Single & none \\
      Case~A & during MS \\
      Case~Be & between TAMS and TACHeB, radiative envelope \\
      Case~Bl & between TAMS and TACHeB, convective envelope \\
      Case~C & after TACHeB \\
      \hline
    \end{tabular}
    \tablefoot{Case~A binary stripping or mass accretion takes place during the MS evolution of the donor and accretor, respectively, Case~B between the TAMS and terminal age core helium burning (TACHeB) and Case~C after TACHeB. Early and late Case~B (Be/Bl) are stars with a radiative and mostly convective envelope, respectively.}
  \end{center}  
\end{table}

\subsection{1D CCSN explosion model}
\label{sec:sn-m}

The \citetalias{M16} model was used to predict the outcome of iron core collapse in all the stellar models used in this work. \citetalias{M16}  takes as the input the density, chemical composition, binding energy, sound speed, and entropy profiles from the SN progenitor. 
The CCSN outcomes predicted by the \citetalias{M16} code depend on in total six explicitly specifiable free parameters. The accretion efficiency $\zeta = 0.8$, the cooling timescale $\tau_\mathrm{1.5}/s = 1.2$ of a $1.5 \, M_\odot$ PNS and the mass outflow fraction $\alpha_\mathrm{out} = 0.5$ are left with the default values from \cite{M16}. For the \citetalias{M16} model used in this work, a lower shock compression ratio value, $\beta = 3.3$ (default: $4.0$), along with a greater shock expansion parameter due to turbulent stresses, $\alpha_\mathrm{turb} = 1.22$ (default: $1.18$) is adopted, and the maximal gravitational NS mass is lowered to $M_\mathrm{NS, grav}^\mathrm{max} = 2.0 \, M_\odot$ (default: $2.05 \, M_\odot$).
In \cite{Schneider21}, these parameters were tuned with the goal to obtain an average explosion energy of Type~IIP SNe in the range of $0.5-1.0 \, \mathrm{B}$ for consistency with observations. 
For single stars, this choice of parameters preserves the highly skewed shape of the explosion energy distribution landscape, which is also obtained with the default parameters; however, it admits explosion energies above $2 \, \mathrm{B}$, which is the maximal value over the H16 progenitors when exploded with the \citetalias{M16} model with default parameters. The explosion energy distribution from our customized parameter choice has a longer tail; namely, an extension toward greater explosion energies up to $ \simeq 3 \, \mathrm{B}$ from single-star progenitors\footnote{These high explosion energies represent a regime that has not been tested against observations and multi--D simulations.} at $Z=Z_\odot$ (see Fig. \ref{fig:fallback-mco}).  
The predicted chirp-mass landscape of binary black hole (BH) mergers obtained with this parameter choice is consistent with the LIGO-Virgo-KAGRA observations of gravitational wave sources after the third observing run \citep{Schneider2023}.

\subsection{Catalog of 3D CCSN simulation outcomes}
\label{sec:3d-models}
Our explodability criteria emulate the predictions of the semi-analytical \citetalias{M16} code, which has been developed in the Monash group. The 3D CCSN modeling approaches in the Monash and Garching groups are comparable; the main difference is the approximated neutrino transport in the Monash group, which reduces the computational costs of the simulations. Therefore, we compare the final fate predictions made using our pre-SN criteria to those of state-of-the-art 3D CCSN simulations from the archives of the Garching and Monash groups.
The pre-SN properties of the progenitors and the CCSN outcomes are summarized in Tables \ref{Tab:garching} and \ref{Tab:monash} in the appendix. 

Multidimensional (Multi-D) effects enhance shock revival due to turbulent stresses, the increased advection timescale and the increased heating efficiency compared to the axisymmetric treatment \citep{Mueller2015}. 
One approach to effectively model the shock revival enhancing effects in 1D is to modify the equations for hydrostatic structure and for the jump conditions at the shock by scaling up the shock radius, $r_\mathrm{sh} \, {\rightarrow} \, \alpha_\mathrm{turb} \cdot r_\mathrm{sh}$, by a factor $\alpha_\mathrm{turb}$ that is set by the root-mean-square averaged turbulent Mach number in the gain region \citep{Mueller2015}. Physically, one of the origins for strong seed perturbations are oxygen-neon shell mergers \citep{Mueller2019}, convective burning in the silicon burning phase or pulsations before the iron core surpasses its effective Chandrasekhar mass. \cite{M16} suggested 
that explosion-enhancing multi-D effects are switched off at $\alpha_\mathrm{turb} \, {\simeq} \, 0.86,$ rather than at $\alpha_\mathrm{turb} \, {=} \, 1$ due to re-normalisation procedures. 
Since in our parameter choice for \citetalias{M16}, we kept the value $\alpha_\mathrm{turb} \, {=} \, 1.22$ fixed, we implicitly assume the presence of a seed perturbation or other shock-revival enhancing multi-D effects (such as magnetic fields). Therefore, in our comparison of SN progenitor properties to outcomes of 3D CCSN simulations, we use those carried out with seed perturbations or magnetic fields whenever available. 

We compile 3D CCSN simulation outcomes for eight different single-star progenitors obtained in the Garching group and 21 different single-star and binary-stripped star SN progenitors from the Monash group. These simulations have been carried out either over spherically symmetric progenitors with or without magnetic fields or with 3D progenitor stratifications as initial conditions, obtained, for instance, from simulations of precollapse core oxygen burning. 

\subsection{Supervised learning model}
\label{sec:sl-m}

The usual way to predict CCSN outcomes with \citetalias{M16}
is to evolve stellar models up to the onset of iron-core infall and to use the pre-SN structure profiles as input. However, since the stellar structure profiles are not available beyond the stellar parameter grid nodes over which massive stars have been evolved up to the pre-SN stage, it is not possible to directly predict final fates 
over a quasi-continuous parameter space as necessary for BPS studies. 
To overcome this gap, we directly relate the stellar parameters to CCSN outcomes. We achieve this by fitting the scalar quantities used to formulate the explodability criteria as a function of the progenitor $M_\mathrm{CO}$, while taking into account differences between single and binary-stripped stars and the $Z$ dependence.   
As a fitting model, we use a Gaussian process regression \citep[GPR;][]{Rasmussen2004}. 
Its kernel is a free choice that needs to be preset before training the model using the supervised learning method.
In this work, we adapt the kernel choice to each fitting task. We often found the best performance with the Matérn kernel.
The Matérn kernel has the free parameters $\nu$ and $l$: while $\nu$ controls the smoothness of the approximated function,
$l$ is the length scale of the kernel. The training task is to optimize the choice of the kernel parameters by returning a probability distribution over their values, so that the resulting multivariate normal has maximum likelihood over the training data. 

Once the GPR model is fitted, point predictions are made by conditioning over the training data and the prediction intervals are obtained from marginalization. More details are given in Sect.~\ref{app:explod_prox}. 

\begin{figure*}
      \centering 
      \includegraphics[width=0.45\textwidth]{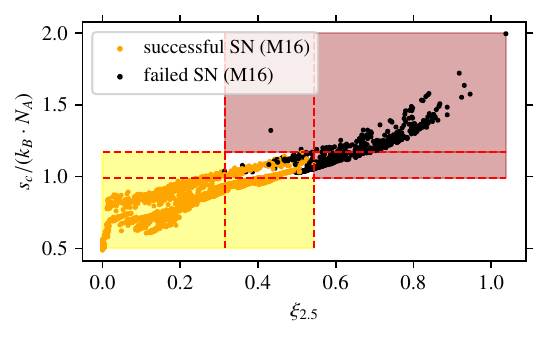}
      \includegraphics[width=0.45\textwidth]{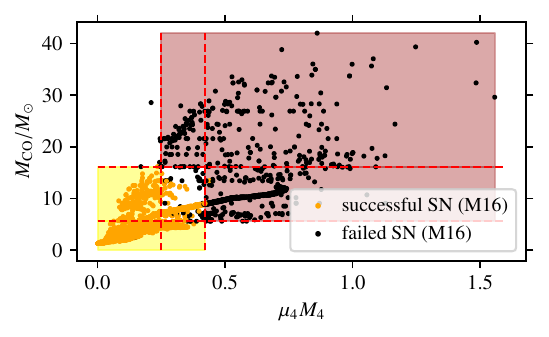}
      \caption{
       Final fate assignment in the $(\xi_\mathrm{2.5},s_c)$ and $(\mu_4 M_4, M_\mathrm{CO})$ planes, based on CCSN outcomes predicted by \citetalias{M16} over the set of \citetalias{Schneider21}, \citetalias{Schneider2023}, \citetalias{Schneider2024}, \citetalias{Temaj}, and H16 SN progenitors. The red lines indicate the critical lower and upper thresholds of each explodability proxy considered, which are summarized in Table~\ref{Tab:proxies}. The background color classifies the explodability based on these thresholds: a failed SN region (brown), a successful SN region (yellow) and unclassified, when using these two variables (blank). The fates of those collapsing stars that hitherto have not been classified are mapped out by a separation line in the $(\mu_4M_4, \mu_4)$ plane (see right panel of Fig.~\ref{fig:ertl}).}
       \label{fig:proxies}
    
\end{figure*}

\begin{figure*}
  \centering
  \includegraphics[width=0.45\textwidth]{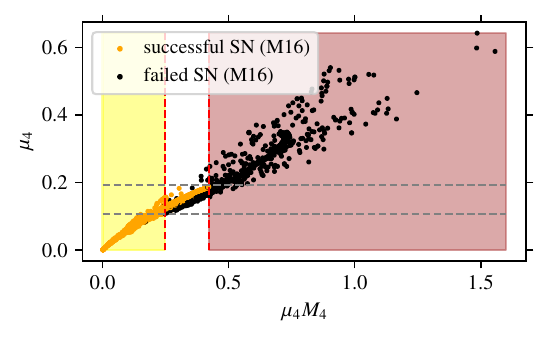}
  \includegraphics[width=0.45\textwidth]{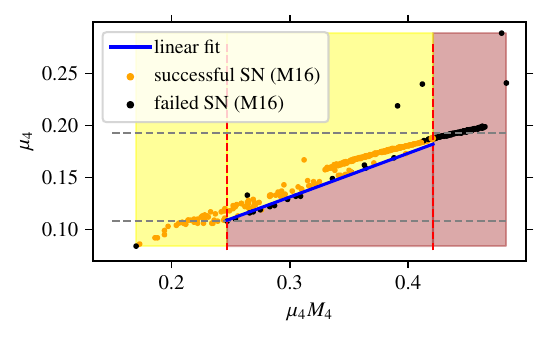}
  \caption{Final fate distribution in the $(\mu_4 M_4, \mu_4)$ plane over the entire set of \citetalias{Schneider21}, \citetalias{Schneider2023}, \citetalias{Schneider2024}, \citetalias{Temaj}, and H16 SN progenitors (left panel) and over a subset of progenitors that fall into the ``overlap region'' with degenerate final fate outcomes, given that that their $\xi_\mathrm{2.5}$, $s_c$, and $M_\mathrm{CO}$ values are neither above nor below the upper or lower threshold value, respectively, for assignment of a final fate forecast based on these variables alone (right panel). Stars from this ``overlap region'' explode if $\mu_4 M_4$ is below the lower threshold. They fail to explode if $\mu_4 M_4$ is greater than the upper threshold. For $\mu_4 M_4$ values in between, a separation line (in blue) discriminates the CCSN outcomes. The background color classifies the explodability based on this scheme, separating the failed SN (brown) from the successful SN (yellow) regions.}
\label{fig:ertl}
\end{figure*}

\section{Results}

\subsection{Pre-SN explodability criteria}
\label{sec:shock-revival-univ-preSN}

We find that the final fate cannot be discriminated by using a threshold value of a single explodability proxy, whether that is $\xi_\mathrm{2.5}$, $s_c$ or iron core mass, $M_\mathrm{Fe}$, nor by using threshold values of pairs of explodability proxies (e.g., $\xi_\mathrm{2.5}$ and $s_c$), nor by a separation line in the $(\mu_4 M_4, \mu_4)$ plane (see Fig.~\ref{fig:proxies} and \ref{fig:ertl}). 
However, the CCSN outcome is predicted reliably when using multiple diagnostic pre-SN variables and the criteria detailed in the next subsection.

\subsubsection{Successful versus failed SN}
If either $\xi_\mathrm{2.5}$, or $s_c$ or $M_\mathrm{CO}$ are above an upper threshold value, the SN outcome is a failed SN (for the critical values, see Table~\ref{Tab:proxies}). 
If, conversely, any of these proxies is below a lower threshold value, the outcome is a successful SN. For values in the intermediate (``overlap'') region or in the rare case of conflicting final fate assignments using these variables, the final fate is decided in the $(\mu_4 M_4, \mu_4)$-plane. If the $\mu_4 M_4$ coordinate is critically low (large), the outcome is a successful (failed) SN, while for intermediate values a separation line tells the exploding and the non-exploding stars apart. The separation line is set by the parameters $(k_1, k_2)=(0.005, 0.420)$ and yields the following condition for a failed SN, over $\mu_4 M_4 \in (0.247, 0.421)$:
\begin{equation}
    \mu_4 < k_1 + k_2 \cdot \mu_4 M_4.
    \label{Eq:mu4}
\end{equation}
Stellar models that fulfill this condition, namely,\ those found below the separation line in the $(\mu_4 M_4, \mu_4)$ plane, are non-exploding; otherwise, they explode. We note that according to \citetalias{Ertl2016}, the final fate outcomes are reversed: the non-exploding models are found above the separation line. 

\begin{table}
  \begin{center}
  \caption{Upper and lower threshold values in pre-SN stellar structure variables that predetermine CCSN outcomes, as predicted by \citetalias{M16}. 
  }
    \begin{tabular}{l l l}       
    \hline\hline  
       variable $X$ & lower threshold $X^\mathrm{min}$ & upper threshold $X^\mathrm{max}$\\
      \hline
      $ \xi_{2.5}$ & 0.314 & 0.544\\
      $ s_c / [N_\mathrm{A} \cdot k_\mathrm{B}]$ & 0.988 & 1.169 \\
      $ M_\mathrm{CO}/ \mathrm{M_\odot}$ & 5.6 & 16.2 \\
      $ \mu_4 M_4$ & 0.247 & 0.421 \\
      \hline
    \end{tabular}
  \label{Tab:proxies}
  \end{center}  
\end{table}

Out of 3987 progenitor models, 2685 ($\simeq 69 \%$) explode. The pre-SN explodability scheme introduced above
replicates the CCSN outcomes predicted by \citetalias{M16} with an accuracy of $\simeq 99.4$ \%; therefore, it can be used reliably as its surrogate. 

The main idea advocated by our approach is that
the explodability proxies $\xi_\mathrm{2.5}$, $s_c$ and the \citetalias{Ertl2016} parameters $(\mu_4 M_4, \mu_4)$ are not equivalent in their significance for the final fate outcome. While these can in some cases follow similar trends (e.g., a large $\xi_\mathrm{2.5}$ accompanied by a large $s_c$), they may carry complementary information about explodability (e.g., a critically low $s_c$ for a successful SN, but a moderate $\xi_\mathrm{2.5}$ within the degenerate range over which failed and successful SNe coexist).
To assign the final fate at the pre-SN stage, insight from several explodability proxies needs to be drawn together. Our formalism states that it is sufficient to evaluate $\xi_\mathrm{2.5}$, $s_c$, $\mu_4$, $\mu_4 M_4$, and $M_\mathrm{CO}$. These proxies probe the SN progenitor at four different mass coordinates: 1) the innermost region of the stellar core at constant entropy, 2) the mass coordinate at which  the PNS is typically enclosed, 3) at $2.5 \, M_\odot$, and 4) at the carbon-oxygen-rich layers much further out.

\subsubsection{Compact remnant after a successful SN}
\label{sec:fallback}

\begin{figure*}
      \centering
      \includegraphics[width=0.45\textwidth]{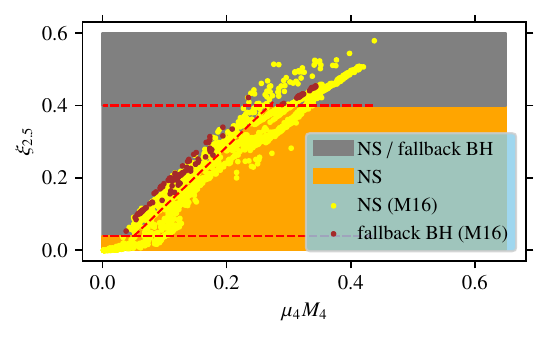}
      \includegraphics[width=0.45\textwidth]{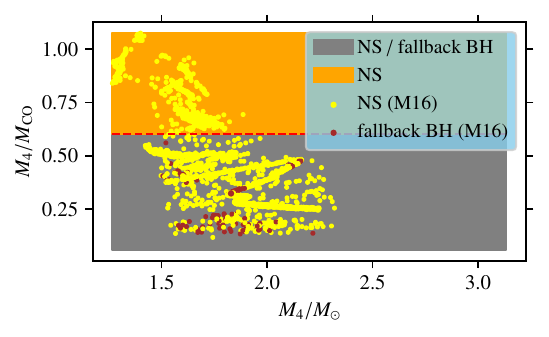}
      \caption{Prerequisites for BH formation by fallback in a successful SN explosion.       
      Left panel:  Compact remnant type left behind the explosion is constrained in the $(\mu_4M_4, \xi_\mathrm{2.5})$ plane. The lower boundary in $\xi_\mathrm{2.5}$, demarcating the region over which only NSs are obtained, is indicated by a red horizontal line. The upper boundary in $\xi_\mathrm{2.5}$, above which NS and fallback BH remnants coexist, is also indicated by a red horizontal line. For $\xi_\mathrm{2.5}$ values in between, the linear $\xi_\mathrm{2.5}$-to-$\mu_4 M_4$ fit model (represented by a diagonal dashed red line) sets the minimal $\xi_\mathrm{2.5}$ value for fallback to occur depending on $\mu_4M_4$.       
      Right panel:  Compact remnant type is also constrained by the ratio $M_4 / M_\mathrm{CO}$. The $M_4 / M_\mathrm{CO} < 0.6$ condition for fallback to occur is represented by a red horizontal line through the $(M_4, M_4/M_\mathrm{CO})$ plane.      
      In both panels, the parameter space region that permits the coexistence of NSs and fallback BHs is shown in grey, while the region that constraints the remnant to be a NS is shown in orange. }    
    \label{fig:pre-SNfb}
\end{figure*}

The  \citetalias{M16} model not only predicts the final fate (successful or failed SN) but also the remnant type (NS or fallback BH) for the case of a successful SN. After shock revival and SN launch, the mass accretion onto the PNS  continues while the mass is ejected. 
In the \citetalias{M16} model, a BH forms by fallback, if the predicted initial explosion energy is insufficient to unbind the envelope or if the mass accreting PNS surpasses the maximal equation-of-state dependent baryonic NS mass, $M_\mathrm{NS, bary}^\mathrm{max}$. Otherwise, the compact remnant is predicted to be a NS.
Physically, there are at least three different channels for explosive BH formation:
\begin{enumerate}
    \item  delayed collapse of the hot PNS after neutrino cooling, due to losses in thermal pressure contributions to its stability \citep{Baumgarte1996},
    \item  collapse of the PNS through the accretion of initially ejected fallback matter, falling back if its radial velocity eventually decreases below the escape velocity \citep{Colgate1971} and
    \item  collapse of the PNS through the accretion of infall matter from the progenitor core, preceded by successful shock revival and affecting the subsequent explosion dynamics \citep{Chan2018}.
\end{enumerate}
While explosive BH formation via the first and second channels happens on
timescales of tens of seconds to hours after shock revival \citep{Wong2014}, the third channel leads to BH formation on much shorter timescales, hundreds of milliseconds after shock revival \citep{Chan2018, Chan2020}.
The \citetalias{M16} model takes into account only the second explosive BH formation channel, and therefore likely underestimates systematically the occurrence of BH formation in successful SNe.

In the following, we use the terms ``direct BH'' and ``fallback BH.'' Direct BH formation (within the context  and terminology applied in this paper) envisions the intermediary formation of a PNS that collapses after accretion pushes its mass above the stability limit before shock revival could take place (i.e., via a failed SN).
This discrimination allows us to decouple the question of a star's compact remnant type from that of its compact remnant mass, which is an additional degree of freedom associated with uncertainties.\footnote{A scheme to predict the compact remnant mass inevitably needs to make assumptions about what fraction of the hydrogen-rich envelope (if any is left) falls into the BH formed by direct collapse, and what fraction of the ejecta mass falls back onto the PNS when a fallback BH is formed. Such a scheme is not the subject of this work, but see \cite{Boccioli} and \cite{ugolini2025}  for recent compact remnant mass fitting models based on $\xi_M$ and on $M_\mathrm{CO}$, respectively.}

Out of 2685 exploding SN progenitors in our sample, only 167 ($\simeq 6 \%$) leave behind fallback BHs.
Based on pre-SN properties alone, we do not find a scheme that could discriminate the remnant type (NS versus fallback BH) deterministically (however, see Sect.~\ref{app:det_fb} for an approach based on the explosion energies). Instead, we empirically identify prerequisites placed on pre-SN variables that need to be fulfilled for fallback BH formation to occur with a specific probability:
\begin{itemize}
    \item Fallback BH formation occurs only if the compactness is not critically low. If
    \begin{equation}
        \xi_\mathrm{2.5} < 0.04,
    \end{equation}
    then only NSs are left behind, since the silicon-oxygen layers (typically found at the $2.5 \, M_\odot$ mass coordinate surrounding the infalling iron core) are then too loosely bound for substantial fallback onto the PNS.  
    \item Fallback BH formation may occur if the dimensionless $\xi_\mathrm{2.5}$ is large compared to the dimensionless $\mu_4M_4$. 
    We find that only NSs are left behind if
    \begin{equation}
        \xi_\mathrm{2.5} < a \cdot \mu_4 M_4 + b
    ,\end{equation}
    with $(a,b)= (1.75, -0.044)$. This condition remains valid for $\xi_\mathrm{2.5} \in (0.04, 0.4)$. 
    \item If
    \begin{equation}
        \xi_\mathrm{2.5} > 0.4,
    \end{equation}
    the outer layers surrounding the iron core typically are tightly bound in our models and fallback may occur regardless of how large $\mu_4 M_4$ is in comparison.
    \item Fallback BH formation is found only if $M_\mathrm{CO}$ is large compared to the PNS mass (which correlates with the $M_4$ coordinate) which controls the neutrino heating via the PNS surface temperature \citep{Mandel2020}. 
    If
    \begin{equation}
        M_4 / M_\mathrm{CO} > 0.6,
    \end{equation}
    only NSs are left behind in our models.
\end{itemize}

In Fig. \ref{fig:pre-SNfb}, we summarize these prerequisites for BH formation by fallback. We find that among the exploding progenitors that satisfy these conditions, the fraction of SN explosions that lead to fallback BHs is $f=0.15$, while we obtain only NS formation if these conditions are not satisfied. 

\subsection{Final fate landscapes of single and binary-stripped stars}
\label{sec:ff_land}
We now use the explodability and fallback criteria based on the pre-SN structure of the SN progenitors, presented in Sect.~\ref{sec:shock-revival-univ-preSN},  to assign final fates (Sect.~\ref{sec:sys-dep-rev}) and compact remnant types (Sect.~\ref{sec:ns_windows}) to a set of single star and binary-stripped star models at variable metallicity. We present a CCSN recipe applicable for rapid BPS that derives from these (Sect.~\ref{sec:met}) and generalize our findings (Sect.~\ref{sec:xc}).

\subsubsection{Explodability dependence on $M_\mathrm{CO}$ and $Z$}
\label{sec:sys-dep-rev}

There are three main reasons why it is preferable to parametrize explodability as a function of $M_\mathrm{CO}$ rather than $M_\mathrm{ZAMS}$. First, $M_\mathrm{CO}$ together with $X_C$ -- the central carbon mass fraction at the end of core-helium burning -- controls the core evolution through the late burning phases and thereby  pre-determines the pre-SN stellar structure and explodability (see Sect.~\ref{sec:disc}). Second, $M_\mathrm{CO}$ provides a more direct link than $M_\mathrm{ZAMS}$ to SN observations and thus avoids additional sources of systematic uncertainty (see Sect.~\ref{sec:comp_obs}). Third, $M_\mathrm{CO}$ typically remains nearly constant from the time of formation at the end of core-helium burning up to iron-core collapse, while the relation between $M_\mathrm{ZAMS}$ and $M_\mathrm{CO}$ depends on the entire evolutionary history of the star and binarity effects. While the relation between $M_\mathrm{CO}$ and $M_\mathrm{ZAMS}$ is close to linear for single stars, the slope differs substantially depending on the stellar evolution model (see Fig.~\ref{fig:heger-schneider} for an example), and these masses can therefore not be used interchangeably.  Fig.~\ref{fig:heger-schneider} also shows that the $M_\mathrm{ZAMS}$--to--$M_\mathrm{CO}$ relation for binary-stripped stars becomes nonlinear when $Z$ varies.

Previous works have shown that for the same $M_\mathrm{CO}$, pre-SN profile structures of massive stars differ, depending on metallicity \citep[e.g.,][]{LC2018, Schneider2023} and on whether the star evolved in isolation or underwent binary mass transfer \citep[e.g.,][]{Brown2001, Wellstein1999, Schneider21, Schneider2024, Laplace21}. In spite of this, $M_\mathrm{CO}$--based CCSN recipes that are typically used in rapid BPS codes do not make such distinctions. 

Here, our goal is to predict the outcome of CCSNe as a function of $M_\mathrm{CO}$ 
in a class-specific way, namely,\ with a distinct treatment of single and binary-stripped stars, and by taking $Z$--dependence into account. To this end, we restrict the models to those of \citetalias{Schneider21} and \citetalias{Schneider2023} to have a homogeneous set of models computed with the same code and physics assumptions.

For $M_\mathrm{CO} > M_\mathrm{CO}^\mathrm{max}$ and regardless of the values of the other explodability criteria, the final fate is a failed SN. For $M_\mathrm{CO} < M_\mathrm{CO}^\mathrm{min}$, the final fate is a successful SN.
Within the range $M_\mathrm{CO} \in (M_\mathrm{CO}^\mathrm{min}, M_\mathrm{CO}^\mathrm{max})$ that has coexistence of failed and successful SN outcomes, we sample -- for each $Z$ and MT history class of the star -- $M_\mathrm{CO}$ in increments of $\delta M_\mathrm{CO}/M_\odot = 0.1$  and evaluate the fitted GPR models of $\xi_\mathrm{2.5}$, $s_c$, $\mu_4 M_4$ and $\mu_4$. 
If any of the $\xi_\mathrm{2.5}$, $s_c$ or $\mu_4 M_4$ is above (below) the upper (lower) threshold determined in Sect.~\ref{sec:shock-revival-univ-preSN}, 
the outcome at the corresponding $M_\mathrm{CO}$ is a failed (successful) SN. If not yet classified given these threshold values, the $\mu_4$ fit model is used to discriminate the final fate outcome based on the separation line in the $(\mu_4M_4, \mu_4)$ plane set by Eq.~(\ref{Eq:mu4}). 

Not only the explodability proxies such as $\xi_\mathrm{2.5}$ and $s_c$ of single and binary-stripped stars \citep[e.g.,][]{Sukhbold2018, LC2018, PS20, Schneider21, Schneider2023, Takahashi,Laplace2025}, but all the explodability proxies relevant for our criteria introduced in Sect.~\ref{sec:shock-revival-univ-preSN} follow structured bimodal trends with $M_\mathrm{CO}$, characterized globally by two peaks and a valley in-between (see Fig.~\ref{fig:proxies-fit}). Tables~\ref{Tab:mco-failed-succZ} and~\ref{Tab:mco-failed-succZ10} summarize the predicted final fate outcomes of single and binary-stripped stars at the two metallicities $Z=Z_\odot$ and $Z=Z_\odot/10$.

\begin{table}[h!]
  \begin{center}
  \caption{Boundary values in $M_\mathrm{CO}$ demarcating non-explosive BH formation by direct collapse at $Z = Z_\odot$. We expect the direct BH outcomes for $M_\mathrm{CO} > M_\mathrm{CO}^{(3)}$ to extend up to the pair-instability mass gap.
  }
    \begin{tabular}{l l l l}       
    \hline\hline       
       & $M_\mathrm{CO}^{(1)}/M_\odot$ & $M_\mathrm{CO}^{(2)}/M_\odot$ & $M_\mathrm{CO}^{(3)}/M_\odot$   \\
      \hline
      Single & 6.6 & 7.2 & 13.0\\
      Case~C & 6.6 & 7.1 & 13.2 \\
      Case~Bl & 7.7 & 8.3 & 15.2 \\
      Case~Be & 7.8 & 8.3 & 15.3\\
      Case~A & 7.4 & 8.4 & 15.4 \\
      \hline
    \end{tabular}
  \label{Tab:mco-failed-succZ}
  \end{center}  
\end{table}  

\begin{table}[h!]
  \begin{center}
  \caption{Same as Table \ref{Tab:mco-failed-succZ}, but for $Z = Z_\odot/10$. } 
    \begin{tabular}{l l l l}       
    \hline\hline       
       & $M_\mathrm{CO}^{(1)}/M_\odot$ & $M_\mathrm{CO}^{(2)}/M_\odot$ & $M_\mathrm{CO}^{(3)}/M_\odot$   \\
      \hline
      Single & 6.1 & 6.6 & 12.9\\
      Case~C & 6.3 & 7.1 & 12.3 \\
      Case~Bl & 7.0 & 7.9 & 14.0 \\
      Case~Be & 6.9 & 7.9 & 13.5\\
      Case~A & 6.9 & 7.4 & 13.7 \\
      \hline
    \end{tabular}
  \label{Tab:mco-failed-succZ10}
  \end{center}  
\end{table} 

The following structural pattern of explodability dependence on $M_\mathrm{CO}$ is preserved, regardless of $Z$ and binarity: 
\begin{itemize}
    \item $M_\mathrm{CO} < M_\mathrm{CO}^{(1)}$: successful SNe;    \item $M_\mathrm{CO}^{(1)} \leq M_\mathrm{CO} \leq M_\mathrm{CO}^{(2)}$: window of BH formation by direct collapse;
    \item $M_\mathrm{CO}^{(2)} < M_\mathrm{CO}<  M_\mathrm{CO}^{(3)}$: successful SNe;
    \item $M_\mathrm{CO} \geq M_\mathrm{CO}^{(3)}$: BH formation by direct collapse.
\end{itemize}
Differences in $Z$ and binary MT history change the boundaries $M_\mathrm{CO}^{(1)}$, $M_\mathrm{CO}^{(2)}$, and $M_\mathrm{CO}^{(3)}$ -- but not the general pattern. Comparing the final fates of single and binary-stripped stars at $Z=Z_\odot$ and at $Z=Z_\odot/10$ over the $M_\mathrm{CO}$ range (see Table~\ref{Tab:mco-failed-succZ}, Table~\ref{Tab:mco-failed-succZ10}, and Fig.~\ref{fig:proxies-fit}), the two most important conclusions are as follows: 
\begin{enumerate}
    \item The boundaries $M_\mathrm{CO}^\mathrm{(1)}$, $M_\mathrm{CO}^\mathrm{(2)}$ and $M_\mathrm{CO}^{(3)}$ shift systematically toward lower values as $Z$ decreases from $Z_\odot$ to $Z_\odot / 10$. Here,  $M_\mathrm{CO}^{(3)}$ is greater, the earlier the hydrogen-rich envelope is removed. 
    \item The critical $M_\mathrm{CO}$ values of single stars and Case~C donors are similar, and differ more substantially from those of Case~A and B donors, which are also similar. The need to discriminate between single and stripped star SN progenitors is apparent. For example, at $Z = Z_\odot$, the BH formation windows by direct collapse of the Case~C and Case~A donors do not even overlap.
\end{enumerate}

\subsubsection{NS formation from explosions of $M_\mathrm{CO} > 7 \, M_\odot$ cores}
\label{sec:ns_windows}
After having mapped out the final fates of single and binary-stripped star models in Sect.~\ref{sec:sys-dep-rev}, as a next step, we investigate the compact remnant type left behind after a successful SN, which is either a NS or a fallback BH. We inquire whether the rarer fallback BH formation outcome can be excluded based on progenitor $M_\mathrm{CO}$, $Z,$ and MT history class.

Evaluating the conditions for BH formation by fallback defined in Sect.~\ref{sec:fallback} based on the pre-SN variables $\xi_\mathrm{2.5}, \mu_4M_4, M_4$, and $M_\mathrm{CO}$, we find ranges in $M_\mathrm{CO}$ over which these are not satisfied; namely, ranges over which the compact remnant is guaranteed to be a NS.
While the variables $\xi_\mathrm{2.5}$, $\mu_4 M_4$, and $M_4 = \mu_4 M_4 / \mu_4$ all show bimodal trends with $M_\mathrm{CO}$ and sharply decrease for $M_\mathrm{CO} > M_\mathrm{CO}^{(2)}$, it is the difference in the slopes at which these quantities decrease (increase) compared to one another that constrains the remnant type. 

Tables~\ref{Tab:ns-wind} and \ref{Tab:ns-wind2} summarize the widest intervals $\delta M_\mathrm{CO}$ 
for which we obtain exclusively NS formation at $Z=Z_\odot$ and at $Z = Z_\odot/10$, over the Case~A, Case~B\footnote{The pre-SN properties of Case~Be and Case~Bl donors are similar, and the distinction between Case ~Bl and Case~Be systems is not always trivial in the context of rapid BPS studies. We thus coarse-grain the fits of $\xi_\mathrm{2.5}$, $\mu_4M_4$ and $M_4$ necessary for evaluation of our probabilistic fallback model over the Case~Be and Case~Bl donors at $Z_\odot$ and at $Z_\odot / 10$, respectively.}, Case~C, and single-star SN progenitors within the range of $M_\mathrm{CO}^\mathrm{(2)} < M_\mathrm{CO} < M_\mathrm{CO}^{(3)}$. The \citetalias{M16} model predicts the formation of NSs from the SN explosions of massive progenitors with $M_\mathrm{CO} > 7 \, M_\odot$, which CCSN recipes such as those introduced in \cite{Fryer2012} and \cite{Mandel2020} do not allow for.

In our sample of the \citetalias{Schneider21} and \citetalias{Schneider2023} stellar models, we only find SN progenitors that leave behind fallback BHs for $M_\mathrm{CO}^{(2)} < M_\mathrm{CO} < M_\mathrm{CO}^{(3)}$ (see the right panel in Fig.~\ref{fig:single-stripped}).
Stellar models that fulfill the criteria defined in Sect.~\ref{sec:fallback} form fallback BHs with a frequency of 0.15, which we interpret as a probability $P=0.15$. The probability has an objective and a subjective origin: First, we expect that the map from $M_\mathrm{CO}$ to the remnant type is only partially bijective (i.e.,\ allowing for the co-existence of NSs and fallback BHs over bands in $M_\mathrm{CO}$; see Sect.~\ref{app:h16_ff} for support of this assumption). Second, we are ignorant of the precise location and width of the window over which fallback BHs are expected to cluster.

\begin{table}[h!]
  \begin{center}
  \caption{\label{Tab:ns-wind} Critical values in $M_\mathrm{CO}$ for NS formation at $Z = Z_\odot$ from single- and binary-stripped star SN 
  progenitors}. 
    \begin{tabular}{l l l}       
    \hline \hline       
       & $M_\mathrm{CO}^\mathrm{NS, 1}/M_\odot$ & $M_\mathrm{CO}^\mathrm{NS, 2}/M_\odot$ \\
      \hline
      Single & 9 & 10.2 \\
      Case~C & 9.6 & 10.7 \\
      Case~B & 9.9 & 10.3 \\
      Case~A & 11.1 & 12.1  \\
      \hline
    \end{tabular}
    \tablefoot{These values indicate the width of the windows for $ M_\mathrm{CO}\in (M_\mathrm{CO}^\mathrm{NS,1}, M_\mathrm{CO}^\mathrm{NS, 2})$ over which single stars, Case~C, Case~B, and Case~A stripped stars are expected to explode and to leave behind NSs, respectively. Outside these windows, the compact remnant for exploding stars with $M_\mathrm{CO}^{(2)} < M_\mathrm{CO} < M_\mathrm{CO}^{(3)}$ is either a NS or a fallback BH.} 
  
  \end{center}  
\end{table}  

\begin{table}[h!]
  \begin{center}
  \caption{Same as Table \ref{Tab:ns-wind}, but at $Z = Z_\odot / 10$.} 
    \begin{tabular}{l l l}       
    \hline \hline      
       & $M_\mathrm{CO}^\mathrm{NS, 1}/M_\odot$ & $M_\mathrm{CO}^\mathrm{NS, 2}/M_\odot$ \\
      \hline
      Single & 7.4 & 11 \\
      Case~C & 8.9 & 9.5 \\
      Case~B & 9.3 & 10.3 \\
      Case~A & 10.4 & 11.1  \\
      \hline
    \end{tabular}
  \label{Tab:ns-wind2}
  \end{center}  
\end{table}  

We observe the following structural pattern: for the same MT history class, the window in $M_\mathrm{CO}$ with NS formation guaranteed shifts systematically toward larger values as $Z$ increases.

\subsubsection{CCSN recipe for rapid BPS}
\label{sec:met}

\begin{figure*}
      \centering
      \includegraphics[width=0.41\textwidth]{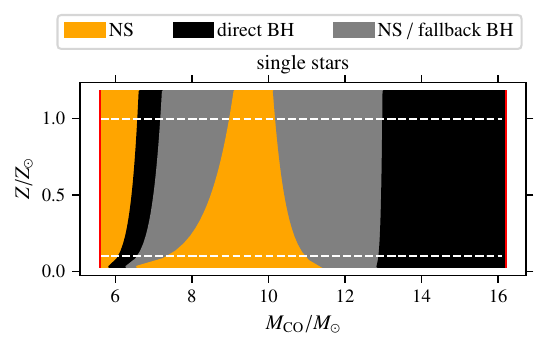}
      \includegraphics[width=0.41\textwidth]{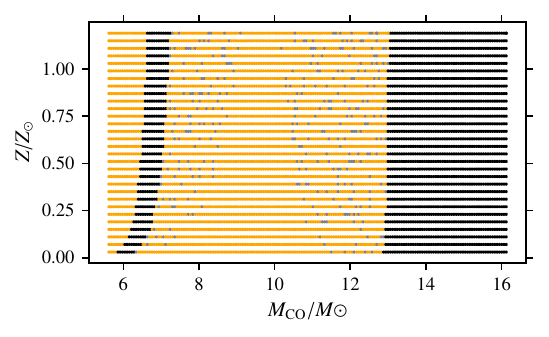}
      \includegraphics[width=0.41\textwidth]{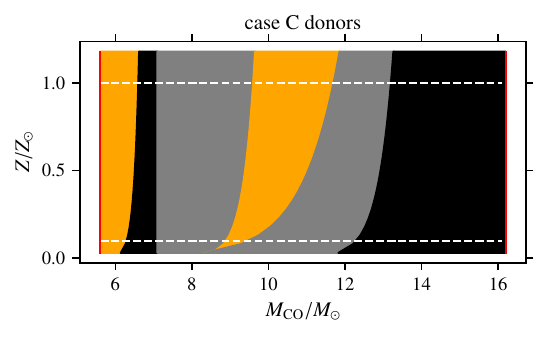}
      \includegraphics[width=0.41\textwidth]{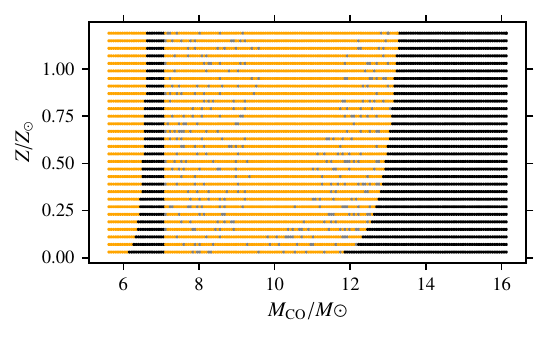}
      \includegraphics[width=0.41\textwidth]{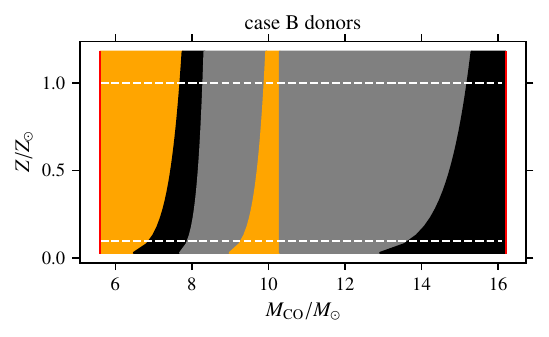}
      \includegraphics[width=0.41\textwidth]{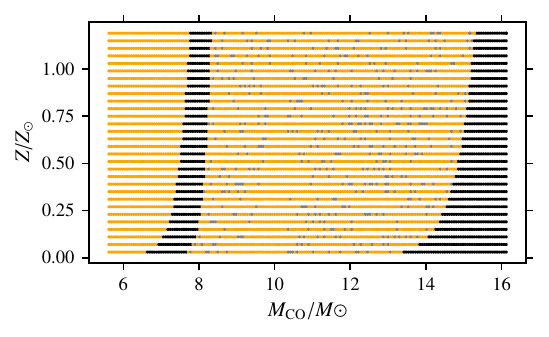}
      \includegraphics[width=0.41\textwidth]{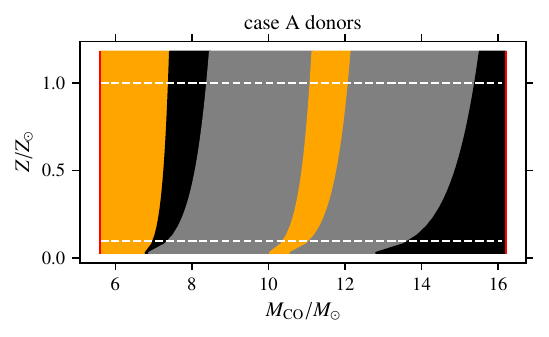}
      \includegraphics[width=0.41\textwidth]{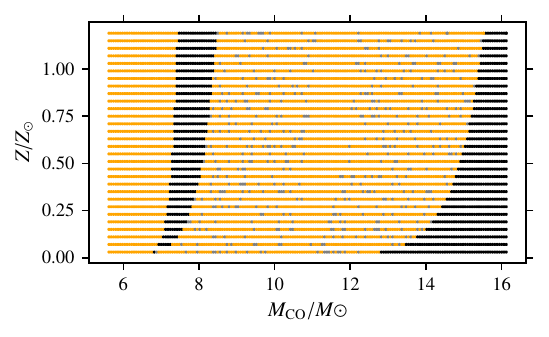}
      \caption{Left panels: Regions of BH formation by direct collapse (in black), of NS formation (in orange) and of the coexistence of fallback BH and NS remnants (in grey) in the stellar parameter space spanned by $M_\mathrm{CO}$ and $Z$ for each MT history class (single stars, Case~C, Case~B, and Case~A donors), as predicted by the CCSN recipe introduced in this work. The $M_\mathrm{CO}$ range is limited to $M_\mathrm{CO} \in (M_\mathrm{CO}^\mathrm{min}, M_\mathrm{CO}^\mathrm{max})$ (borderlines in red), outside of which only successful and failed SNe, respectively, are predicted to occur. For $M_\mathrm{CO} < M_\mathrm{CO}^\mathrm{(1)}$, NS formation is guaranteed. For $M_\mathrm{CO} \in (M_\mathrm{CO}^{(1)}, M_\mathrm{CO}^{(2)})$, the CCSN outcomes are failed SNe, leaving behind BHs formed by direct collapse. For $M_\mathrm{CO} \in (M_\mathrm{CO}^\mathrm{(2)}, M_\mathrm{CO}^\mathrm{(3)})$, successful SNe are guaranteed. Over this range, the compact remnant is a NS if $M_\mathrm{CO} \in (M_\mathrm{CO}^\mathrm{NS, 1}, M_\mathrm{CO}^\mathrm{NS, 2})$. Otherwise, the remnant is a NS, at a probability of $P=0.85$, or a fallback BH, at a probability of $P=0.15$. The $Z$--dependence is modeled using $M_\mathrm{CO}^\mathrm{(i)} \propto \log (Z/Z_\odot)$.   
      Right panels: Example of a statistical realization of the CCSN recipe with probabilistic fallback BH formation.}       
    \label{fig:mco-windows}
\end{figure*}

Integrating the quantitative results from Sect.~\ref{sec:sys-dep-rev} and \ref{sec:ns_windows}, we construct a CCSN recipe that retains the distinction between single stars and binary-stripped stars and a dependence on $Z$. 

To get a first-order estimation of how BH formation boundaries, which we derived 
for $Z_\odot$ and $Z_\odot/10,$ scale with $Z \in (Z_\odot /10, Z_\odot)$, we assume a linear model in  $\log Z$ given by
\begin{equation}
\label{eq:mcocrit-z}
    M_\mathrm{CO}^\mathrm{(i)}(Z)/M_\odot = a_i + b_i \cdot \log Z/Z_\odot
,\end{equation}
for $i=1,2,3$ and each MT history class. The argument for the parametric form $M_\mathrm{CO}^\mathrm{(i)}(Z)/M_\odot \propto \log Z$ is generic; namely, it is that BH formation regimes scale with $\log Z$
rather than linearly with $Z$, due to the effect of stellar mass loss \citep{Heger2003}. The parametric form should be revised when detailed stellar evolution models of single stars and binary-stripped stars for at least a third grid point in $Z$ will be available. 

The linear model $f(x) = a + b \cdot x$ has two free parameters $a$ and $b$, and we use pairs of data points $\left[ M_\mathrm{CO}^\mathrm{(i)}(Z_\odot), Z_\odot \right]$ and $\left[ M_\mathrm{CO}^\mathrm{(i)}(Z_\odot/10), Z_\odot/10 \right]$ given in Tables~\ref{Tab:mco-failed-succZ} and \ref{Tab:mco-failed-succZ10} to determine these analytically for each MT class: the curve for the critical boundary $M_\mathrm{CO}^\mathrm{(i)}(Z)$ is given by the parameters:
\begin{itemize}
    \item $a = M_\mathrm{CO}^\mathrm{(i)}(Z_\odot)$  and
    \item $b = M_\mathrm{CO}^\mathrm{(i)}(Z_\odot) - M_\mathrm{CO}^\mathrm{(i)}(Z_\odot/10)$.
\end{itemize}
The same formalism is applied to the guaranteed NS formation windows that delineate boundaries of non-zero fallback BH formation probability, listed in Tables~\ref{Tab:ns-wind} and \ref{Tab:ns-wind2}.

For $M_\mathrm{CO} < M_\mathrm{CO}^{(1)}$, since we do not encounter BH formation by fallback in our \citetalias{Schneider21} and \citetalias{Schneider2023} samples, we infer that
these are statistically insignificant and -- for the CCSN recipe -- adopt the assumption that only NSs form. This conclusion is also consistent with our deterministic fallback BH formation criterion (see Fig.~\ref{fig:fallback-mco}).

The prediction of the occurrence of fallback is limited by the restricted modeling approach using only global parameters such as $M_\mathrm{CO}$ and $Z$ given at the time of evolutionary cutoff. Physically, stochasticity in the outcomes can be expected to arise due to the effects of turbulence and magnetohydrodynamics during core-collapse.
We therefore construct a second, even simpler probabilistic fallback model for rapid BPS, which we designate as fallback model B. For model B, we assume a uniform probability of 10\% for the occurrence of fallback BH formation in--between $M_\mathrm{CO}^\mathrm{(2)}$ and $M_\mathrm{CO}^\mathrm{(3)}$. This probability assumption is motivated by the relative frequencies of the occurrence of fallback BH formation in successfully exploding stellar models satisfying $M_\mathrm{CO} > M_\mathrm{CO}^\mathrm{(2)}$:
\begin{itemize}
    \item 8.5\% over the entire set of SN progenitors (\citetalias{Schneider21}, \citetalias{Schneider2023}, \citetalias{Schneider2024}, \citetalias{Temaj}, H16);
    \item 11.5\% over the single stars and binary-stripped stars (\citetalias{Schneider21}, \citetalias{Schneider2023}, \citetalias{Temaj}), 
     adopting the same model for the late burning phases.
\end{itemize}
These frequencies are similar despite the differences in adopted stellar evolution physics and resulting pre-SN properties.  Thus, we coarse-grain over these and adopt an average value of 10\%.

Figure~\ref{fig:mco-windows} shows the predicted compact remnant type landscapes for single stars and binary-stripped stars, which follows from the bimodal dependence of explodability proxies $\mu_4 M_4$, $\mu_4$, $\xi_\mathrm{2.5}$, and $s_c$ on $M_\mathrm{CO}$ (see Fig.~\ref{fig:proxies-fit}).

Finally, with the fitted parameters of the assumed scaling law given by Eq. (\ref{eq:mcocrit-z}), we obtain a CCSN recipe applicable for rapid BPS studies, predicting the final fate (failed or successful SN) and the remnant type (NS or BH) for given $M_\mathrm{CO}$, $Z$, and MT history class (see Sect.~\ref{app:ccsn_rec} for further remarks on its usage). 

\begin{figure*}
      \centering
      \includegraphics[width=0.4\textwidth]{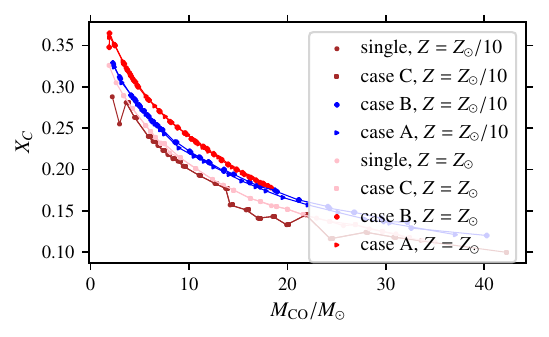}
      \includegraphics[width=0.4\textwidth]{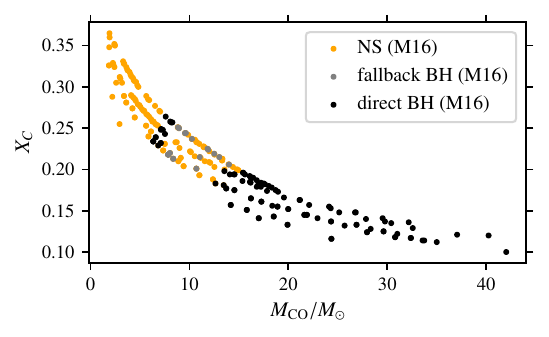}
      \caption{Left panel: $X_C(M_\mathrm{CO})$--tracks of single and binary-stripped stars (Case~C, Case~B, and Case~A donors) at $Z_\odot$ and at $Z_\odot/10$, respectively, from the catalogs \citetalias{Schneider21} and \citetalias{Schneider2023}. Right panel: Their CCSN outcomes, as predicted by \citetalias{M16}.} 
    \label{fig:single-stripped}
\end{figure*}

\begin{figure*}
      \centering
      \includegraphics[width=0.45\textwidth]{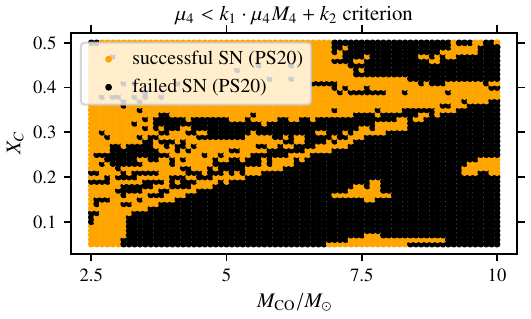}
      \includegraphics[width=0.45\textwidth]{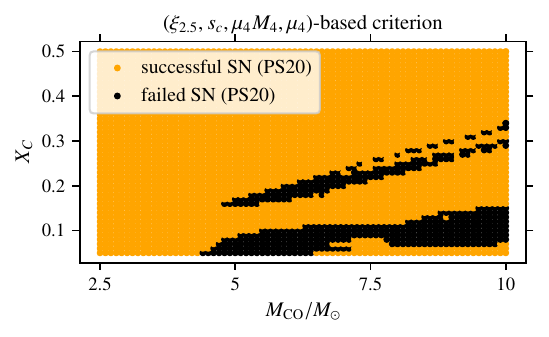}
      \caption{Final fate landscapes resulting from \citetalias{Ertl2016} (left panel) and from our explodability criteria (right panel), when applied to the same pre-SN core profiles over the densely sampled \cite{PS20} grid. In their work, bare CO core have been evolved up to the onset of iron-core infall from different starting points in the $(M_\mathrm{CO}, X_C)$ plane at core-carbon ignition. 
      For the \citetalias{Ertl2016} criterion, the parameters $(k_1, k_2)$ of the separation line are calibrated to the updated W20 model from \cite{Ertl2020}.} 
    \label{fig:ps20-s2123}
\end{figure*} 

\subsubsection{Explodability dependence on $M_\mathrm{CO}$ and $X_C$}
\label{sec:xc}

For the same SN progenitor CO core mass, $M_\mathrm{CO}$, and the same model for the late burning phases, we find systematics behind the differences in final fate outcomes depending on progenitor $Z$ and MT history. These trace back to differences in the respective values of the central carbon mass fraction, $X_C$, at the end of core-helium burning (CHeB), as detailed below.  

For the same $M_\mathrm{CO}$ and MT history, as $Z$ increases, $X_C$ increases (see Fig. \ref{fig:single-stripped}). This is because with greater $Z$, the helium core grows less massive due to its reaction to stronger wind mass loss from the hydrogen-rich envelope. Similarly, in binary-stripped stars of the same $Z$, earlier removal of the hydrogen-rich envelope also leads to a lower helium core mass, as a response to the mass loss into MT. 

In a less massive helium core, the core temperature is lower and the $^{12}\mathrm{C}\left(\alpha, \gamma\right)^{16}\mathrm{O}$ reaction sets in later during core-helium burning. This leaves more carbon in the core at core-helium exhaustion. With a higher $X_C$ for the same $M_\mathrm{CO}$, more nuclear fuel is available during the relatively long-lasting core-carbon burning phase.
A higher $X_C$ shifts the peaks in explodability proxies such as $\xi_\mathrm{2.5}$ toward larger $M_\mathrm{CO}$ values while for decreasing $X_C$, the structural dependence of explodability proxies on $M_\mathrm{CO}$ flattens until the peak structure vanishes altogether \citep{PS20}. The shift of the failed SN window in $M_\mathrm{CO}$ toward larger values as $X_C$ increases is clearly apparent in the right panels of Fig.~\ref{fig:single-stripped} and Fig.~\ref{fig:ps20-s2123}. The physical effects of variable $M_\mathrm{CO}$ and $X_C$ onto the pre-SN structure -- and thereby its final fate -- through the advanced burning phases are discussed in Sect.~\ref{sec:disc}.

Single stars and Case~C donors follow similar tracks in the $(M_\mathrm{CO}, X_C)$ plane at $Z_\odot$ and at $Z_\odot/10$, respectively (see the left panel of Fig.~\ref{fig:single-stripped}). The same applies to the $X_C(M_\mathrm{CO})$ tracks of Case~A and Case~B donors. 
For the same $M_\mathrm{CO}$ and $Z$, Case~A+B donors have a higher $X_C$ compared to single stars + Case~C donors. The Case~A+B MT has a stronger effect on the core structure of the donor than Case~C MT because before core--carbon burning, the core and envelope evolution are not yet decoupled from one another. 

More generally, differences in the adopted stellar evolution physics up to the end of CHeB manifest themselves in differences in $X_C(M_\mathrm{CO})$ relations \citep[e.g.,][]{ChieffiLimongi2020, Schneider21, Temaj}, which in turn take effects on the final fates. 
For example,  we observe that the shifts in the $\xi_\mathrm{2.5}$ peaks toward lower $M_\mathrm{CO}$ values in the H16 compared to the \citetalias{Schneider21} single star models can be traced back to a lower $X_C$ for the same $M_\mathrm{CO}$ in H16 compared to \citetalias{Schneider21} (see Fig.~\ref{fig:heger-schneider-comp2}). 

While we relate the differences in CCSN outcomes between single stars and binary-stripped stars and with variable $Z$ to a higher $X_C$ in binary-stripped stars compared to single stars, and to a higher $X_C$ with increasing $Z$, rapid BPS codes do not track of the $X_C$ variable. BPS codes that do keep track the $X_C$ variable, such as \textsc{posydon} \citep{fragos} and \textsc{bpass} \citep{Byrne}, typically use the densely sampled \cite{PS20} grid of models of core evolution through the late burning phases to look up the final fate. To the final profiles,  \citetalias{Ertl2016} is applied as the default SN model (see left panel of Fig.~\ref{fig:ps20-s2123} and Sect.~\ref{sec:ccsn_recipes} for details) in \textsc{posydon} (see also \cite{Patton2022} for an application in \textsc{bpass}). To make final fate predictions 
with our explodability scheme as an alternative SN model, the $(\xi_\mathrm{2.5}, s_c, \mu_4M_4, \mu_4)$ variables necessary for its evaluation are readily available in the \cite{PS20} data base and can be interpolated over in the $(M_\mathrm{CO}, X_C)$ plane for $M_\mathrm{CO} \leq 10 \, M_\odot$ in the same way as is done at present with the $(\mu_4 M_4, \mu_4)$ parameters for evaluating the \citetalias{Ertl2016} SN model. The difference in the resulting final fate landscape is dramatic (see Fig.~\ref{fig:ps20-s2123}) and further analysed in Sect.~\ref{sec:explo_prox}. 

\subsection{Comparison with other CCSN models}
\label{sec:comp_3d}

We first validate our explodability criteria with the help of 3D CCSN simulation outcomes (Sect.~\ref{sec:3D_ccsn}), and then compare them to other criteria based on $\xi_\mathrm{2.5}$ and \citetalias{Ertl2016} (Sect.~\ref{sec:explo_prox}). Similarly, we compare our CCSN recipe to other $M_\mathrm{CO}$-based SN models and benchmark these based on 3D CCSN simulation outcomes in Sect.~\ref{sec:comp_rec}.

\subsubsection{3D CCSN simulation outcomes}
\label{sec:3D_ccsn}

\begin{figure*}
      \centering
      \includegraphics[width=0.32\textwidth]{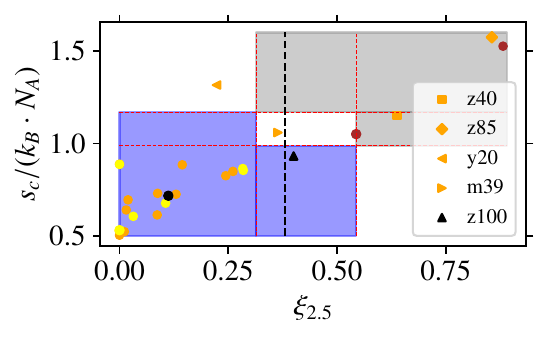}
      \includegraphics[width=0.32\textwidth]{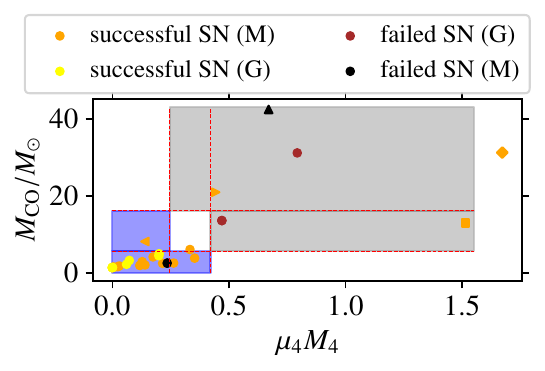}
      \includegraphics[width=0.32\textwidth]{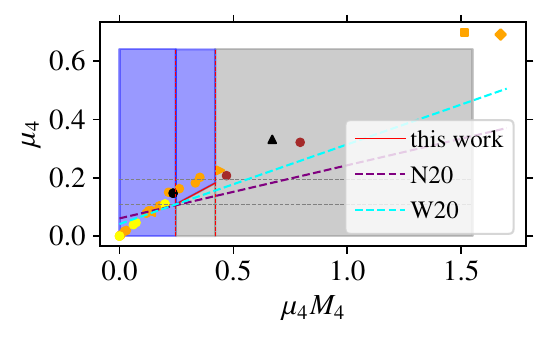}
      \caption{Comparison of the final fate predictions using our pre-SN explodability criteria to 3D CCSN simulation outcomes performed in the Garching (G) and Monash (M) groups. 
      The dashed red lines in the two-parameter planes spanned by $(\xi_\mathrm{2.5}, s_c)$ and by $(\mu_4 M_4, M_\mathrm{CO})$ -- in the left and the central panel, respectively -- indicate the lower and upper thresholds of each explodability proxy used in our pre-SN criteria. In the left panel, the dashed black line shows the explodability classification based on the $\xi_\mathrm{2.5}^\mathrm{crit}=0.38$ threshold.
      In the right panel, the separation line for the final fate classification in the $(\mu_4 M_4, \mu_4)$ plane by the reversed \citetalias{Ertl2016} criterion introduced in this work is shown by a red line, and compared to the standard \citetalias{Ertl2016} criterion, with calibrations from \cite{Ertl2020} for the updated W20 and N20 models (dashed lines in cyan and purple).
      The 3D CCSN simulation outcomes are color-coded in yellow (Garching) and orange (Monash) for the exploding progenitors, and in brown (Garching) and black (Monash) for the non-exploding progenitors. Specific progenitors that are referenced in the main text are represented by symbols other than circles. The background color shows the final fate assignment using our explodability proxies: failed SN (in grey), successful SN (in blue) and unclassified (blank) regions of their value spaces.} 
    \label{fig:3d}
\end{figure*}

For the validation of our pre-SN explodability scheme based on 3D CCSN simulation outcomes, we take into account 6 exploding and 2 non--exploding models from the Garching group archive, as well as 19 exploding and 2 non-exploding models from that of the Monash group. Fig.~\ref{fig:3d} compares the outcomes of 3D CCSN simulations to the predictions made using our explodability criteria that only evaluate the SN progenitor properties. The number of false final fate assignments from applying our explodability scheme to the set of 8+21=29 progenitors is 4, yielding an overall accuracy of 86\% over 3D CCSN simulation outcomes. 

We first analyze the models that explode in 3D. As with the \citetalias{M16} outcomes, most of them are scattered in the region of critically low $\xi_\mathrm{2.5}$ and critically low $s_c$. There are a few exceptions:

\begin{itemize}
    \item the stripped-star model y20 has a conflicting final fate prediction issued by $\xi_\mathrm{2.5}$ (to explode) versus by $s_c$ (to not explode). Its $M_\mathrm{CO} \simeq 8.2 \, M_\odot$ is within the overlap region of degenerate final fate outcomes. It is predicted to explode -- in agreement with the 3D outcome -- due to a critically low $\mu_4 M_4$.
    \item The stripped-star model m39 is within the overlap region in the $(\xi_\mathrm{2.5}, s_c)$ plane, but is predicted to not explode because of a critically large $M_\mathrm{CO} \simeq 21 \, M_\odot$, in disagreement with the 3D outcome. 
    Its $\mu_4 M_4 = 0.443 > (\mu_4 M_4)^\mathrm{max} = 0.421$ is close to the edge case, where the final fate is decided by the separation line.    
    \item The Population III star models z85 and z40 having an $M_\mathrm{CO}$ of around 31 and 13 solar masses, respectively, have  critically large $\xi_\mathrm{2.5}$ and $\mu_4 M_4$ for a failed SN  according to our pre-SN criteria, but explode nevertheless in the 3D simulations. Their $(\mu_4 M_4, \mu_4)$ values reach beyond the region sampled by our set of stellar models outlined in Sect.~\ref{sec:prog-m}. We suspect that the reason for the discrepancy in the final fate outcome predictions are effects of the altered nuclear burning and evolution in Population~III stars leading up to iron core collapse compared to stars initially with metals. We conclude that our pre-SN criteria break down when it comes to Population~III stars, which were not part of the set of stellar models based on which these were formulated. We aim to introduce possible fixes to our pre-SN criteria for stellar progenitors with $(\mu_4 M_4, \mu_4)$--values beyond the region sampled by our models and for better applicability to Population~III stars in future work. 
    \end{itemize} 

Three out of the four prediction errors committed by the pre-SN criteria compared to the 3D outcomes are false negatives for an explosion, i.e.\ the 3D simulations seem to be even more optimistic about explosions than our explodability scheme.
It should be noted that the model z40 was specifically triggered for obtaining an explosion, that z85 explodes as a pair-instability SN and that the rapidly rotating m39 was designed as a long gamma-ray burst precursor. \citetalias{M16} does not take rotation effects during collapse into account and is a model for the neutrino-driven perturbation-aided SN engine. 

All models that are non-exploding in the 3D simulation are predicted to not explode by the pre-SN criteria, except for the spherically symmetric progenitor model s14 from the Monash group archive. Whether or not s14 explodes in a 3D simulation if asymmetry is introduced into the progenitor stratifications has not been tested.\footnote{Interestingly, s14 is predicted to result in a failed SN by \citetalias{M16} when setting $\alpha_\mathrm{turb} = 0.86$ while keeping the rest of the parameters as stated in Sect.~\ref{sec:sn-m}. As summarized in Sect.~\ref{sec:3d-models}, setting $\alpha_\mathrm{turb} = 0.86$ is a means to ``switch off'' the shock revival enhancing effects due to turbulent stresses.} The final fate predictions for the non-exploding models with the pre-SN criteria have the following origins: 
\begin{itemize}
    \item The u75 model (represented by a brown circle) has critically large $\xi_\mathrm{2.5}$, $s_c$, $M_\mathrm{CO}$, and $\mu_4 M_4$ values for a failed SN.
    \item The s40 model (represented by a brown circle) has its $\xi_\mathrm{2.5}$, $s_c$, and $M_\mathrm{CO}$ values  the overlap region, but it is predicted to result in a failed SN given its critically large $\mu_4M_4$.
    \item The z100 model has a conflicting final fate prediction based on $s_c$ (explosion) and $M_\mathrm{CO}$ (failed SN), but its $\mu_4M_4$ is critically large for a failed SN.
\end{itemize}

When using our pre-SN criteria for predicting CCSN outcomes, the value of the $\mu_4M_4$ variable is found to play the decisive role for the explosion of y20 and for the non-explosion of s40 and z100.  We aim to test the explodability scheme introduced in this work against multi-D simulation outcomes performed in other groups in follow-up work. These comparisons may show less agreement with our formalism. At present, there is no consensus among the multi-D CCSN modeling communities about which SN progenitors explode and which do not; for example, we refer to Fig.~2 in \cite{2025janka}, highlighting differences in the 2D CCSN simulation outcomes between the Garching and the Princeton groups over the same set of SN progenitors. However, the Monash models tend to reflect some patterns also seen by the Princeton group, namely, a relatively robust explosion even at high masses and one BH formation case from a relatively low mass progenitor.

\subsubsection{The $\xi_\mathrm{2.5}$ and ($\mu_4 M_4, \mu_4$) based explodability criteria}
\label{sec:explo_prox}

We inquire what predictive accuracy over the 3D CCSN simulation outcomes can be achieved with the $\xi_\mathrm{2.5}$-- and \citetalias{Ertl2016}-- based explodability criteria.

An even higher accuracy of 90\% (three errors) is achieved by setting a critical $\xi_\mathrm{2.5}^\mathrm{crit} \, {\simeq} \, 0.38 \pm 0.01$ 
for failed SNe (see left panel of Fig.~\ref{fig:3d}). Compared to the \citetalias{M16} model outcomes, the $\xi_\mathrm{2.5}^\mathrm{crit} > 0.38$ criterion predicts substantially more failed SN outcomes (see Fig. \ref{fig:proxies}) and agrees with \citetalias{M16} in only 86\% of the cases over the sample of 3897 SN progenitors considered in this work.

For many stars, the mass coordinate $2.5 \, M_\odot$ is large enough to be located outside the iron core $M_\mathrm{Fe}$ and small enough to be within the mass accreting region surrounding it. Depending on the SN progenitor model, it is often found at the interface between the silicon core and the silicon-enriched oxygen layers \citep{Sukhbold2014}. However, Fig. ~\ref{fig:comp-mfe} illustrates that for a subset of CCSN progenitors, the $2.5 \, M_\odot$ mass coordinate is found inside the iron core.

As is apparent from the right panel of Fig.~\ref{fig:3d}, our explodability formalism is considerably more optimistic about successful explosions than \citetalias{Ertl2016}.
No single separation line with non-exploding models above and exploding models below the line can be drawn to segregate the final fate outcomes using the standard \citetalias{Ertl2016}. 
The updated \textsc{W20} and \textsc{N20} models\footnote{In 1D CCSN simulations with \textsc{prometheus-hotbath}, the excision of the PNS core introduces free model parameters that regulate the neutrino-emission evolution and settling of the hot accretion mantle above the PNS. These are constrained to reproduce the explosion energy, nickel mass, total neutrino energy loss and duration of the neutrino signal of SN 1987A. However, different parameter choices satisfy these observational constraints, and CCSN outcomes vary depending on their calibration. In turn, different model parameters result in different $(k_1, k_2)$ fit parameters \citep{Ertl2016}.} from \cite{Ertl2020} are consistent with explosions at low $\mu_4M_4$ and low $\mu_4$  but evidently overpredict BH formation compared to the 3D outcomes. The \citetalias{Ertl2016} criterion neither is compatible with the distribution of final fate outcomes predicted by \citetalias{M16} over our set of SN progenitors (see left panel of Fig.~\ref{fig:ertl}).

To highlight the differences between our
and the \citetalias{Ertl2016} pre-SN criteria, we compute the resulting final fate landscapes over the same grid of SN progenitors from \cite{PS20} (see Fig.~\ref{fig:ps20-s2123}). 
When the pre-SN models are assigned a final fate using the \citetalias{Ertl2016} criterion, the final fate is sensitive to the starting point in the $(M_\mathrm{CO}, X_C)$ plane, which features a landscape that has explosion islands in failed SN-dominated regions  and vice versa. In contrast, the explodability scheme introduced in this work leads to a segmented final fate landscape, which features two islands of direct BH formation over a similar $M_\mathrm{CO}$ range but different value ranges in $X_C$. The evolutionary tracks of single and binary--stripped stars typically pass through the upper island (see Fig.~\ref{fig:missingRSG_mcoxc}). 

With \citetalias{Ertl2016}, a failed SN-dominated final fate landscape at the grid boundary $M_\mathrm{CO} = 10 \, M_\odot$ is obtained for most $X_C$ values. In contrast, with our explodability criteria, the upper island of direct BH formation decays at $M_\mathrm{CO} \simeq 9 M_\odot$: at such and greater CO core masses, the final fate landscape is dominated by explosions for $X_C > 0.15$. Thus, our explodability formalism implies that the parameter space width in $M_\mathrm{CO}$ in \cite{PS20} is not wide enough to predict at which $M_\mathrm{CO} > 10 \, M_\odot$ (if at all) the final fate landscape becomes failed SN- dominated regardless of $X_C$. For the \citetalias{Schneider21} and \citetalias{Schneider2023} single and binary stripped stars, 
the failed SN-dominated region is reached at $M_\mathrm{CO} > 15.4 \, M_\odot$; however, this threshold is valid only for $X_C$ roughly within 0.1 and 0.2 (see Fig.~\ref{fig:single-stripped}). When taking into account all the SN progenitors compiled in this work as listed in Sect.~\ref{sec:prog-m}, we find that for obtaining  failed SNe only, a CO core mass threshold as high as $M_\mathrm{CO} \geq M_\mathrm{CO}^\mathrm{max} = 16.2 \, M_\odot$ is required (see Table~\ref{Tab:proxies}).

\begin{figure*}
      \centering
      \includegraphics[width=0.85\textwidth]{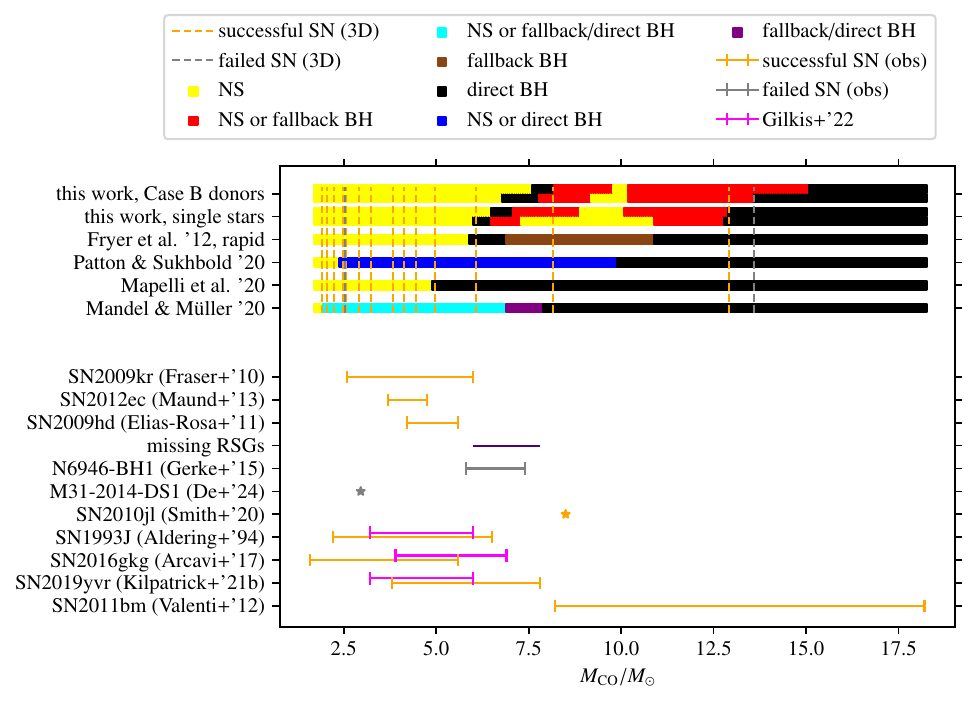}
      \caption{Comparison of $M_\mathrm{CO}$--based CCSN recipes for binary population synthesis (\citetalias{Fryer2012}, \citetalias{Mandel2020}, \citetalias{Mapelli2020}, \citetalias{PS20}, and ours) with 3D CCSN simulation outcomes (from the archives of the Garching and Monash groups) and with SN observations. The compact remnant type as predicted by the recipes -- either a NS, a fallback BH, or a direct BH -- is color-coded as per the legend. For our recipe, we show the predictions for single stars and Case~B donors, respectively, where for each progenitor type the upper stripe evaluates final fates at $Z=Z_\odot$ while the lower stripe those at $Z=Z_\odot / 10$.
      The 3D CCSN simulation outcomes are shown only for progenitors with $2 \leq M_\mathrm{CO}/M_\odot \leq 18$.
      The constraints on the $M_\mathrm{CO}$ range of the progenitors of the three most luminous Type~IIP SNe (SN2009kr, SN2012ec and SN2009hd), of the failed SN candidates (N6946-BH1 and M31-2014-DS1), of the Type~IIn SN (SN2010jl), and of the three most luminous Type~IIb and Ib SNe (SN1993J, SN2016gkg, and SN2019yvr; including the re-estimation of their $\log L_\mathrm{pre-SN,obs}$ from \cite{GilkisArcavi2022}) inferred in this work are plotted below. We also show our estimate of the $M_\mathrm{CO}$ ranges of the missing ``red supergiant problem'' and of the Type~Ic SN2011bm progenitor, when assuming that its remnant is a $1.2 \, M_\odot$ NS.} 
    \label{fig:summary}
\end{figure*}

\subsubsection{$M_\mathrm{CO}$--based CCSN recipes}
\label{sec:comp_rec}

The following, more general conclusions can be drawn from the final fate landscapes of single and binary stripped stars at $Z \geq Z_\odot / 10$ that result from our predictive framework globally, regardless of $Z$ and MT history:
\begin{itemize}
    \item for $M_\mathrm{CO}/M_\odot < 6.1$, only NSs form;    \item for $M_\mathrm{CO}/M_\odot \in (6.1, 15.4)$, NS, direct BH, and fallback BH remnants coexist; 
    \item for $M_\mathrm{CO}/M_\odot \in (8.4, 12.4)$, direct BH formation is excluded and successful SN explosions guaranteed, leaving behind NSs or fallback BHs; 
    \item  for $M_\mathrm{CO}/M_\odot > 15.4$, only direct BHs form.
\end{itemize}
As a reminder, these compact remnant types are not predicted to coexist over the full range in $M_\mathrm{CO}$ for single and binary-stripped stars alike, but rather when the $Z$-dependent compact remnant type predictions for each of these progenitor types are stacked together.

This final fate parametrization using $M_\mathrm{CO}$ differs substantially from others that have commonly been used in BPS codes. In Fig.
~\ref{fig:summary}, we compare the CCSN outcomes predicted by our recipe to 
\begin{itemize}
    \item the ``fast-convection'' CCSN model from \cite{Fryer2012}, which we refer to as rapid \citetalias{Fryer2012};
    \item the \citetalias{Ertl2016}--based CCSN ``look-up table'' from \cite{PS20}, which we refer to as \citetalias{PS20};
    \item the $\xi_\mathrm{2.5}$--based CCSN recipe from \cite{Mapelli2020}, which we refer to as \citetalias{Mapelli2020};
    \item the \citetalias{M16}--based CCSN recipe from \cite{Mandel2020}, which we refer to as \citetalias{Mandel2020}.
\end{itemize}
These recipes are summarized and compared to ours in more detail in Sect.~\ref{sec:ccsn_recipes}. Our CCSN recipe is more optimistic about successful SN explosions than the aforementioned previous works, since it guarantees successful SNe over the widest range at the lower CO core mass end (similar to rapid \citetalias{Fryer2012}), and admits successful SNe over the widest CO core mass range at the high-mass end. 

We assess how compatible the $M_\mathrm{CO}$--based CCSN recipes are with the $M_\mathrm{CO}$ values of progenitors (within the range $1.9 < M_\mathrm{CO}/M_\odot < 16$) that are exploding and non-exploding in the 3D CCSN simulations, respectively. The results are shown in Fig.~\ref{fig:summary}.
All recipes admit explosions for $M_\mathrm{CO} < 5 \, M_\odot$ and, therefore, are consistent with the corresponding 3D CCSN simulation outcomes. The failed SN outcome of the s14 model ($M_\mathrm{CO} = 2.53 \, M_\odot$) is consistent with \citetalias{PS20} and \citetalias{Mandel2020}, but not with the other recipes. The explosion of the s24 single star model ($M_\mathrm{CO} = 6.07 \, M_\odot$, $Z=Z_\odot$) is consistent with our recipe, \citetalias{PS20} and \citetalias{Mandel2020}, but not with the rapid \citetalias{Fryer2012} model or \citetalias{Mapelli2020}. The explosion of the high $M_\mathrm{CO} \simeq 8.2 \, M_\odot$ but low $\xi_\mathrm{2.5} = 0.22$ binary-stripped star y20 is consistent with our recipe, with \citetalias{PS20} and with the rapid \citetalias{Fryer2012} model. It is not consistent with the default upper mass limit for explosions ($M_4^* = 8 \, M_\odot$) in \citetalias{Mandel2020} and with the $\xi_\mathrm{2.5}$-to-$M_\mathrm{CO}$ relation assumed in \citetalias{Mapelli2020}.  The explosion of the Population~III star z40 having $M_\mathrm{CO} = 12.92 \, M_\odot$ is not consistent with any of the CCSN recipes mentioned, though our recipe is closest to allowing for explosions of single star progenitors at such high CO core masses. All CCSN recipes are consistent with the non-explosion of the s40 model having $M_\mathrm{CO} = 13.59 \, M_\odot$ and $Z=Z_\odot$. 
None of them are consistent with the explosions of the $M_\mathrm{CO} \simeq 21 \, M_\odot$ stripped star m39 and of the $M_\mathrm{CO} \simeq 31 \, M_\odot$ Population~III star z85.


\subsection{Comparison with SN observations}
\label{sec:comp_obs}
In what follows, we benchmark the $M_\mathrm{CO}$ based CCSN recipes (\citetalias{Mandel2020}, \citetalias{Mapelli2020}, \citetalias{Fryer2012}, \citetalias{PS20} and ours) based on SN observations that enable  estimates to be made for the $M_\mathrm{CO}$ of the SN progenitors. Our comparison study of CCSN recipes is summarized in Fig.~\ref{fig:summary}, which also synthesizes the comparison with SN observations and with 3D CCSN simulation outcomes.

\subsubsection{Type~IIP SN progenitors and the missing RSG problem}

\begin{figure}
      \centering
      \includegraphics[width=0.45\textwidth]{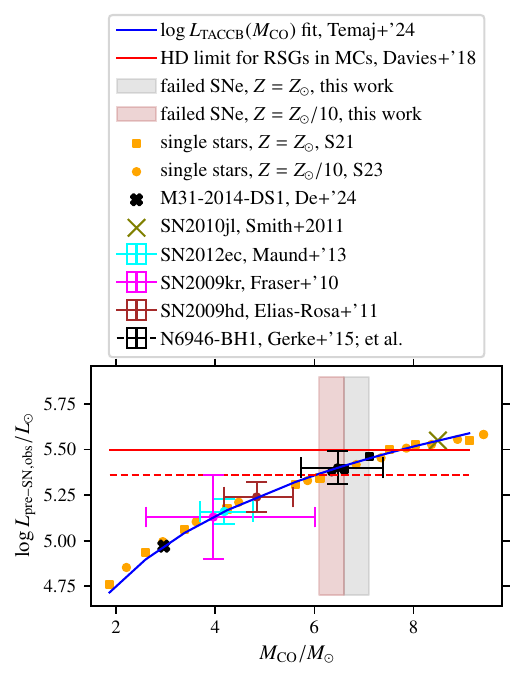}
      \caption{Estimation of the $M_\mathrm{CO}$ values (within uncertainty bounds) of the most luminous progenitors of the Type~IIP SNe SN2012ec, SN2009kr and SN2009hd, of the failed SN candidates N6946-BH1 and M31-2014-DS1 and of the progenitor of the Type~IIn SN2019jl, assuming that all these systems stem from the single star progenitor channel. 
       A parametric scaling law (in blue)
      relates the bolometric luminosity $\log L_\mathrm{TACCB}$ of stellar evolution models (in orange and black) at terminal age core--carbon burning (TACCB) to $M_\mathrm{CO}$. Observations are compared with the failed SN windows in $M_\mathrm{CO}$ (shaded intervals) predicted by our CCSN recipe  as a function of $Z$. The Humphreys--Davidson (HD) limit for RSGs in the Magellanic Clouds and the upper limit on the SN2009kr progenitor luminosity define the value range in $\log L_\mathrm{TACCB}$ (and thus the $M_\mathrm{CO}$ range) of the missing RSGs (red horizontal lines).} 
      \label{fig:missingRSG}
\end{figure}

In a few dozen cases of Type~IIP SN observations in nearby galaxies, the explosion site has been directly imaged years before the transient detection using space- or ground-based telescopes. Pre-explosion imaging allowed to estimate the photometric properties of the 
SN progenitors, in particular their effective temperature and bolometric luminosity, confirming the expectation that these are red supergiants \citep{Smartt2015}. 
These not only preserve a hydrogen-rich envelope up to collapse, but also retain a nearly constant\footnote{The CO core mass changes negligibly from the end of CHeB to the onset of iron-core infall not only for red supergiants (RSGs) but for all the \citetalias{Schneider21} and \citetalias{Schneider2023} SN progenitors (except at $M_\mathrm{CO} > 18 \, M_\odot$).} $M_\mathrm{CO}$ after formation at the end of CHeB. The CO core mass sets the inner temperature and density stratifications, and it thereby determines the burning rate of helium in the shell surrounding the CO core. This leads to a dependence of the bolometric luminosity of pre-SN RSGs, which is mostly sustained by helium shell burning, on $M_\mathrm{CO}$. In the present work, we use the empirical formula given by Eq.~(6) in \cite{Temaj} to estimate\footnote{The 
scaling law from \cite{Temaj} is in agreement not only with the \citetalias{Schneider21} and the \citetalias{Schneider2023} but also the H16 stellar models, despite the differences in adopted evolutionary physics. \cite{Temaj} further showed that it holds regardless of the convective core overshooting assumption. It therefore seems to be a more general model relating the SN progenitor luminosity to $M_\mathrm{CO}$.} $M_\mathrm{CO}$ of observed Type~IIP SN progenitors. To this end, we invert this equation to express the estimated CO core mass $\hat{M}_\mathrm{CO}$,
\begin{equation}
\label{eq:logLmco}
    \hat{M}_\mathrm{CO}/M_\odot = 10^{(\log L_\mathrm{pre-SN, obs}/L_\odot-4.372)/1.268},
\end{equation}
as a function of the observed pre-SN bolometric luminosity $\log L_\mathrm{pre-SN, obs}$. We then apply it according to the details given below.  
According to \cite{DaviesBeasor2018b}, the three most luminous Type~IIP SN progenitors observed are those of SN2009hd \citep{Elias-Rosa2011}, SN2012ec \citep{Maund2013}, and SN2009kr \citep{Fraser2010}. With its $\log L_\mathrm{pre-SN, obs}/L_\odot = 5.24 \pm 0.08$ and Eq.~(\ref{eq:logLmco}), we estimate that the progenitor of SN2009hd has $\hat{M}_\mathrm{CO} = 4.84^{+0.76}_{-0.66} M_\odot$. The Humphreys-Davidson (HD) limit of the most luminous RSGs that have been observed in the Magellanic Clouds is $\log L/L_\odot \simeq 5.5$ \citep{Davies2018}. The lack of observed Type~IIP SN progenitors at luminosities between the most luminous Type~IIP progenitor and the HD limit
defines the Red Supergiant Problem \citep{Smartt2009, Smarttetal2009}. Assuming that the most luminous Type~IIP SN progenitor is set by the upper limit $\log L_\mathrm{pre-SN, obs}^+/L_\odot = 5.36$ on the SN2009kr progenitor, using Eq.~(\ref{eq:logLmco}), we infer that the RSGs miss out over a CO core mass range of at least\footnote{See the Appendix \ref{app:mRSG_L} for a discussion of the upper and lower boundary of the $\log  L_\mathrm{pre-SN,obs}$ range of the missing RSGs that we adopt for this inference.} $6 < M_\mathrm{CO}/M_\odot < 7.8$. 

One\footnote{Other approaches to address the missing RSG problem include pulsation-driven mass loss \citep{dw2022ApJ} of the supergiants, peeling off their outer layers. These stars then continue to evolve as hotter yellow supergiants.} of the proposed solutions is that RSGs over this range do not explode and instead form direct collapse BHs \citep{Smartt2009, Smartt2015}. We test whether this hypothesis is compatible with our predictive models. Since our $M_\mathrm{CO}$--based CCSN recipe for single stars predicts explosions for $M_\mathrm{CO} < 6.1 \, M_\odot$ at $Z > Z_\odot /10$, the three most luminous Type~IIP SNe and observations of all fainter ones are all consistent with our predictive model (see Fig.~\ref{fig:missingRSG}).  Our CCSN recipe predicts failed SNe 
within the value range in $M_\mathrm{CO}$ over which RSGs are indeed found to be missing. However, this range in $M_\mathrm{CO}$ does not explain the missing RSGs over $5.45 \leq \log L_\mathrm{pre-SN, obs}/L_\odot \leq 5.5$.
This means that according to our CCSN recipe, failed SNe can be part of the solution to the missing RSG problem, %
but there must be other physical reasons that explain the lack of Type~IIP SN progenitors in particular over the highest luminosity range.

The source N6946-BH1 is a failed SN candidate \citep{Gerke2015}, whose bolometric luminosity $\log L_\mathrm{pre-SN, obs}/L_\odot = 5.40 \pm 0.09$ imaged before disappearing in the optical \citep{Adams2017} is within the luminosity range of the missing RSGs. From Eq.~(\ref{eq:logLmco}) then follows $\hat{M}_\mathrm{CO} = 6.5^{+0.92}_{-0.73}\,M_\odot$ for its progenitor. Its fate of a failed SN  is consistent with the direct BH formation interval in $M_\mathrm{CO}$ predicted by our CCSN recipe. However, the observation is not constrained enough to confirm it (see Fig.~\ref{fig:missingRSG}). 
An alternative explanation is a stellar merger scenario, discussed, for instance, in \cite{Adams2017} and in \cite{Kashi2017}.
After the recent JWST observations of a luminous infrared source at the same sky location, the interpretation of N6946-BH1 remains ambiguous (see \citealt{Beasor2024} and \citealt{Kochanek2024} for two opposing views).

For the failed SN candidate M31-2014-DS1 \citep{de2024}, the authors infer a much fainter $\log L_\mathrm{pre-SN,obs}/L_\odot = 4.97$. With this estimate, we obtain a progenitor $\hat{M}_\mathrm{CO} = 2.96 \, M_\odot$. A failed SN at such a low CO core mass is incompatible with our model.

\begin{figure}
      \centering
      \includegraphics[width=0.45\textwidth]{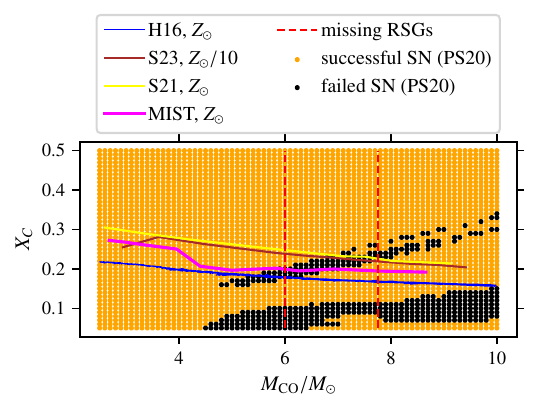}
      \caption{Final fates in the $(M_\mathrm{CO}, X_C)$ plane at the end of CHeB -- obtained by evaluating the explodability scheme introduced in this work over the SN progenitor models from \cite{PS20} -- and the missing RSG problem. To address it via failed SNe, stellar evolution models have to pass through the black-colored region in between the red vertical lines, which delineate the interval in $M_\mathrm{CO}$ over which RSGs are missing as Type~IIP SN progenitors. The H16, \citetalias{Schneider21}, \citetalias{Schneider2023} and \textsc{mesa isochrones and stellar tracks} \citep[MIST;][]{Choi2016} single-star models lead to different $X_C(M_\mathrm{CO})$ tracks through this plane. 
      For any stellar model choice, the missing RSG problem can only partially be addressed, and over a different range in $M_\mathrm{CO}$ (and, thus, also in $\log L_\mathrm{pre-SN,obs}$). 
      } 
    \label{fig:missingRSG_mcoxc}
\end{figure}

We now compare the same observations to the other aforementioned $M_\mathrm{CO}$ based CCSN recipes\footnote{See Sect.~\ref{sec:ccsn_recipes} for a brief summary of the CCSN recipes and for the nomenclature of variables referenced here.}. Since \citetalias{Mandel2020} admits failed and successful SNe up to $M_\mathrm{CO} < 8 \, M_\odot$, it is consistent with observations of the most luminous Type~IIP SN progenitors, however, is not equipped to address the missing RSG problem by failed SNe.
The model \citetalias{Mapelli2020} is in agreement with the brightest Type~IIP SN progenitors.  
The model \citetalias{Fryer2012} is consistent with the most luminous Type~IIP SN progenitors and partially explains the missing RSG problem by BH formation over a similar range 
as our CCSN recipe. The models
\citetalias{Mapelli2020} and \citetalias{Mandel2020} are consistent with N6946-BH1 constituting a failed SN for any value within $\log L_\mathrm{pre-SN,obs}/L_\odot = 5.40 \pm 0.09$, whereas -- similarly to the constraint valid for our CCSN recipe -- \citetalias{Fryer2012} requires it to be within $\log L_\mathrm{pre-SN,obs} / L_\odot = 5.40 \pm 0.04$. The model \citetalias{PS20} explains the most luminous Type~IIP SNe and the missing RSG problem by failed SNe provided that stellar models of RSGs  have suitable $X_C(M_\mathrm{CO})$ relations at the end of CHeB for passing through successful and failed SN sites, respectively (see left panel of Fig.~\ref{fig:ps20-s2123}). When using the same \cite{PS20} catalog of bare CO cores evolved through the late burning stages but our pre-SN explodability criteria instead of \citetalias{Ertl2016} to map out final fates, 
tighter constraints are posed on stellar evolution models at the end of CHeB. In order to ``land'' on the failed SN stripe over $M_\mathrm{CO}$ values in between the missing RSG interval in $M_\mathrm{CO}$, stellar models need to have specific $X_C$ values
over $M_\mathrm{CO}$ intervals (and thereby $\log L_\mathrm{pre-SN}/L_\odot$ value ranges) of interest (see Fig.~\ref{fig:missingRSG_mcoxc}). The only two CCSN recipes that are compatible with the conclusion that the fate of M31-2014-DS1 is a failed SN are \citetalias{PS20} and \citetalias{Mandel2020}.  

\subsubsection{Type~IIn SN progenitors}   
 
In the few cases of SN~IIL progenitor identification, no progenitor luminosity greater than the most luminous Type~IIP of $\log L_\mathrm{pre-SN,obs}/L_\odot = 5.24 \pm 0.08$ has been estimated. In contrast, Type~IIn SN progenitors as bright as $\log L_\mathrm{pre-SN,obs}/L_\odot > 6$ have been observed \citep{Gal-Yam_2007, Boian, kankare2015}.
Type~IIn SNe are distinguished by narrow, bright multi-component hydrogen Balmer lines in the spectrum. These lines are attributed to interactions of the supernova with the circumstellar medium, which may have been formed by episodes of enhanced mass loss from the SN progenitor. 
Progenitors of Type~IIn SNe can be single stars, but they could also be accretor stars and stellar merger products. The latter two categories are the more likely ones, given the progenitor temperatures and luminosities \citep{Justham2014, Schneider2024}. In the accretor star scenario, a binary system is subject to stable mass transfer, wherein the accretor gains mass from the hydrogen-rich envelope of the donor star and then explodes to produce a hydrogen-rich transient. Given an accretor star progenitor, the large $\log L_\mathrm{pre-SN,obs}$ does not necessarily imply a large $M_\mathrm{CO}$, because the bolometric pre-SN luminosity of the blue supergiants is mostly contributed by the hydrogen-rich envelope mass through hydrogen shell burning.

However, not all Type~IIn SNe need to have originated from accretor star or stellar merger progenitors. In the single-star progenitor scenario, the star is expected to have gone through a luminous blue variable (LBV) phase of enhanced mass loss outbursts, which however did not shed  the entire hydrogen-rich envelope by the time of the explosion. We explore consequences of the hypothesis that the Type~IIn SN2010jl \citep{Smith2010} is such a case. Its comparatively faint progenitor is inferred to have a bolometric luminosity of $\log L_\mathrm{pre-SN,obs}/L_\odot = 5.55$, and the photometric data is consistent with a progenitor that has gone through a LBV phase. The $\log L_\mathrm{TACCB}(M_\mathrm{CO})$ scaling law remains reliably applicable to our \citetalias{Schneider21} and \citetalias{Schneider2023} single star models up to $ \log L_\mathrm{TACCB}/L_\odot \simeq 5.7$, even though the stellar models are no longer RSGs at this higher luminosity range.  Thus, using Eq.~(\ref{eq:logLmco}), we estimate a CO core of $\hat{M}_\mathrm{CO} = 8.5 \ M_\odot$ for the SN2010jl progenitor. This value is within the interval in $M_\mathrm{CO}$ over which \citetalias{Fryer2012} and our CCSN recipe predict explosions of single stars and binary-stripped stars independent of $Z$. The \citetalias{Mandel2020} model cannot explain a single star progenitor channel of SN2010jl, so long as it admits explosions only up to $M_\mathrm{CO} < 8 \, M_\odot$. 
The model \citetalias{Mapelli2020} cannot explain the missing RSG problem by failed SNe and SN2010jl by the single star progenitor channel at the same time. Lifting $\xi_\mathrm{2.5}^\mathrm{crit}$ to a greater value to explain the progenitor luminosity of SN2010jl results in loss of explanatory power over the missing RSG problem.

\subsubsection{Type~IIb and Ib SN progenitors}
The progenitors of stripped-envelope SNe (SESNe) are considered to either be massive single stars that experienced strong mass loss shedding away their hydrogen-rich envelopes or donor stars that evolved through a binary MT phase. At the time of the explosion, the progenitor could be a blue supergiant, a cool supergiant, or a Wolf-Rayet (WR) star. 

Only five progenitors of Type~IIb and two progenitors of Type~Ib SNe have been imaged directly \citep{GilkisArcavi2022}. The most luminous progenitors are those of the Type~IIb SN1993J \citep{Aldering}, estimated to have $\log L_\mathrm{pre-SN,obs}/L_\odot = 5.1 \pm 0.3$, of the Type~IIb SN2016gkg \citep{Arcavi}, estimated to have $\log L_\mathrm{pre-SN,obs} / L_\odot = 4.99 \pm 0.32$, and of the Type~Ib SN2019yvr \citep{Kilpatrick21b}, estimated to have $\log L_\mathrm{pre-SN,obs}/L_\odot = 5.3 \pm 0.2$. All three estimates have been revised in \cite{GilkisArcavi2022}, which assesses the most luminous progenitor source to be that of SN2016gkg with $\log L_\mathrm{pre-SN,obs} / L_\odot = 5.28 \pm 0.16$. 

\begin{figure}
      \centering
      \includegraphics[width=0.45\textwidth]{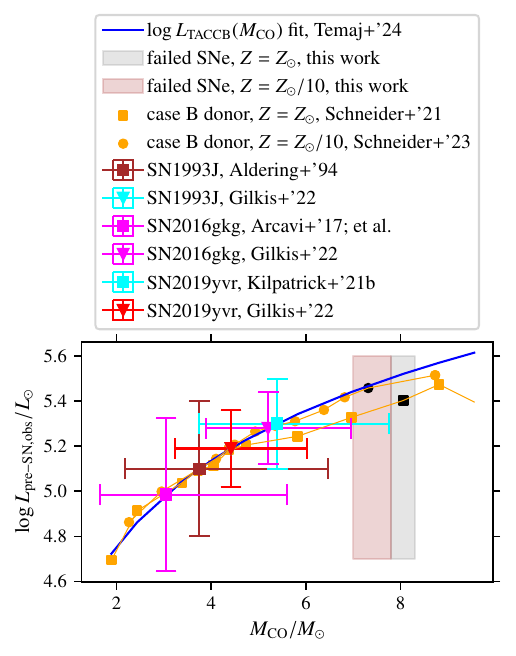}
      \caption{Inferred $M_\mathrm{CO}$ (within uncertainty bounds) of the most luminous Type~IIb and Type~Ib SN progenitors. To obtain the estimates, we use the parametric scaling law $\log L_\mathrm{TACCB}(M_\mathrm{CO})$, which remains reliably applicable -- at any $Z$ -- up to $\log L_\mathrm{TACCB}/L_\odot < 5.2$. For higher SN progenitor luminosities, the scaling law provides a lower limit on $M_\mathrm{CO}$ of the SN progenitor. Observations are compared with the direct BH formation windows for Case~B donors predicted by our CCSN recipe.} 
    \label{fig:TypeIbprog}
\end{figure}

We explore the consequences of the hypothesis asserting that the progenitors of these systems are Case~B donors that explode after having lost all or most of their hydrogen-rich envelope. This progenitor channel is supported by comparison of photometric observations to detailed stellar evolution models \citep{Yoon2017}. 
Within the observational uncertainty bounds, the SESN progenitor $M_\mathrm{CO}$ inferred\footnote{The $\log L_\mathrm{TACCB}(M_\mathrm{CO})$ scaling law is applicable to Case~B donor models from \citetalias{Schneider21} and \citetalias{Schneider2023} independent of $Z$ up to $\log L_\mathrm{TACCB}/L_\odot \simeq 5.2$. At greater $\log L_\mathrm{TACCB}$, dependence on $Z$ emerges: Eq.~(\ref{eq:logLmco}) predicts a $M_\mathrm{CO}$ value lower than the actual stellar models at $Z_\odot/10$ (see Fig.~\ref{fig:TypeIbprog}). Therefore, the scaling law  provides a lower limit on the SESN progenitor $M_\mathrm{CO}$ for $\log L_\mathrm{TACCB}/L_\odot > 5.2$.} from Eq.~(\ref{eq:logLmco}) is admitted to be $> 6 \, M_\odot$ in four out of a total of six aforementioned reference luminosity estimates.
Compared to the most luminous Type~IIP SN progenitors detected, those of Type~IIb and Ib admit greater progenitor luminosities and, therefore, greater $M_\mathrm{CO}$. This trend is consistent with our CCSN recipe, which predicts the BH formation windows for Case~B donors to be shifted toward greater $M_\mathrm{CO}$ values compared to those for single stars. The model from 
\citetalias{Mandel2020} is equipped to explain the progenitor observations, since it allows for explosions at $M_\mathrm{CO} < 8 \, M_\odot$. 
However, the observations challenge the \citetalias{Fryer2012} model, which predicts direct BH formation and no explosions universally for all single and stripped stars satisfying $6 < M_\mathrm{CO}/M_\odot < 7$. Given the large observational uncertainties on the SESN progenitor luminosities, a decisive statement falsifying the rapid \citetalias{Fryer2012} model cannot be made.    

\subsubsection{Type~Ic SN progenitors}
The spectra of Type~Ic SNe lack both hydrogen and helium lines. 
In the single star progenitor channel, these form after an episode of enhanced wind mass loss, such as the WR phase, that removes all or most of the helium-rich envelope. This typically requires a higher $Z$ for strong enough winds. In the stellar binary progenitor channel, a carbon-oxygen star can be formed by helium-rich envelope removal through the combined effect of mass loss by Roche lobe overflow and winds; therefore, it is not limited to higher $Z$. 

Since the helium-rich envelope is then mostly lost by the time the iron core collapses, the final pre-SN mass $M_\mathrm{final}$ cannot be substantially greater than $M_\mathrm{CO}$. If $M_\mathrm{ej}$, the  mass ejected by the Type~Ic SN, can be deduced from the SN light curve, the following simple approach allows us to estimate the progenitor $M_\mathrm{CO}$:
\begin{equation}
\label{eq:mco-mej}
    M_\mathrm{CO} \simeq M_\mathrm{final} = M_\mathrm{rem} + M_\mathrm{ej},
\end{equation}
where $M_\mathrm{rem}$ is the compact remnant mass, for which an assumption needs to be made. We apply this relation to the spectroscopically normal nickel-rich Type~Ic SN 2011bm \citep{Valenti2012}. It poses a challenging test case to CCSN recipes since its ejecta mass is estimated to be $7 \leq M_\mathrm{ej}/M_\odot \leq 17$. 

Our CCSN recipe is consistent with this estimation so long as $7 < M_\mathrm{ej}/M_\odot < 16.6$, if a NS of $M_\mathrm{rem}=1.2 \, M_\odot$ is born and no further constraints are placed on the SN progenitor $Z$ and MT history. This is because our CCSN recipe admits explosions up to $M_\mathrm{CO} = 15.4 \, M_\odot$ at $Z=Z_\odot$ for Case~B donors, and expects NSs to be the more frequent remnant type than fallback BHs. 

The other CCSN recipes are either compatible with this transient over a smaller $M_\mathrm{ej}$ range, or incompatible. The \citetalias{PS20} model admits explosions up to $M_\mathrm{CO} \leq 10 \, M_\odot$ for suitable values of $X_C$ at the end of CHeB. Assuming that the compact remnant is a NS of mass $M_\mathrm{rem} = 1.2 \, M_\odot$, it is consistent with $M_\mathrm{ej} < 8.8 \, \ M_\odot$. In the model \citetalias{Mapelli2020}, $M_\mathrm{CO}^\mathrm{crit} \geq 8.2 \, M_\odot$ is necessary to explain this transient.

The models \citetalias{Mandel2020} and \citetalias{Fryer2012} predict fallback BH remnants for explosions of progenitors with $M_\mathrm{CO} \geq 7 \, M_\odot$. In the \citetalias{Mandel2020} model,  $M_4^* > 9 \, M_\odot$ is required for consistency with this observation, since a minimal stellar-mass BH has $M_\mathrm{rem} > 2 \, M_\odot$.

To satisfy Eq.~(\ref{eq:mco-mej}) with a fallback BH of mass $M_\mathrm{rem} > 2 \, M_\odot$ and $M_\mathrm{ej} \geq 7 \, M_\odot$,  for the progenitor a CO core of mass $M_\mathrm{CO} > 9 \,  M_\odot$ needs to be assumed in the \citetalias{Fryer2012} model, which predicts explosions for $7 < M_\mathrm{CO}/M_\odot < 11$. However a SN explosion of a Type~Ic progenitor with pre-SN mass of $M_\mathrm{final} \simeq M_\mathrm{CO} > 9 \, M_\odot $ leaving behind a fallback BH remnant of mass $M_\mathrm{rem} < M_\mathrm{CO} - 7 \, M_\odot$ is not consistent with the compact remnant mass calculation formalism of the ``fast-convection'' explosion model (Eqs.~16 and 17 in \citetalias{Fryer2012}). It yields  $M_\mathrm{rem} = 7.272 \, M_\odot$ for a SN progenitor of $M_\mathrm{CO} = 9 \, M_\odot$ and $M_\mathrm{rem} = 10.76 \, M_\odot$ for $M_\mathrm{CO} = 10.9 \, M_\odot$, since the fallback mass fraction is predicted to increase with $M_\mathrm{CO}$ in \citetalias{Fryer2012}.

\subsubsection{Supernova remnants}
The CO core masses of SN progenitors can also be constrained by nebular line spectroscopy of SN remnants. After explosive nuclear burning, the ejecta mass of Type~Ic SNe is mostly composed of oxygen and iron-group elements. The nebular line ratio [OI/CaII] is an indicator of the oxygen mass $M_\mathrm{O,ej}$ released during the SN explosion according to the calibrated scaling law,
\begin{equation}
\label{eq:sc}
\log [\mathrm{OI/CaII}] = 0.9 \cdot \log (M_\mathrm{O,ej} / M_\odot) + 0.03
,\end{equation}
which has been inferred\footnote{The link inferred in \cite{Fang2023} between $\log [\mathrm{OI/CaII}]$ and $M_\mathrm{O,ej}$ in Eq.~(\ref{eq:sc}) is established based on a limited set of SN models. Mixing of silicon, calcium or carbon with oxygen can affect the oxygen line cooling and the associated uncertainties have not been estimated.} to hold for SESNe of Type~IIb, Ib, Ic and Ic-BL \citep{Fang2023, Fang2022}. The greatest values among observed Type~IIb and Ib SNe are $\log [\mathrm{OI/CaII}] \simeq 0.5,$ while those among Type~Ic and Ic-BL SNe reach up to $\log [\mathrm{OI/CaII}] \simeq 0.7$ \citep{Taddia2019, Pellegrino2022, Fang2022}. These imply $M_\mathrm{O, ej} \leq 4.29 \, M_\odot$ and $M_\mathrm{O, ej} \leq 6.23 \, M_\odot$, respectively. $M_\mathrm{O, ej}$ takes up a significant fraction $X_\mathrm{O,ej}$ of the total SN ejecta mass,  
\begin{equation}
\label{eq:xoj}
    M_\mathrm{O, ej} = X_\mathrm{O, ej} \cdot M_\mathrm{ej}
,\end{equation}
and $X_\mathrm{O, ej}$ varies with the progenitor core mass.
The mass fraction is found to be $X_\mathrm{O, ej} < 0.5$ for progenitor CO cores up to $M_\mathrm{CO} < 6.6 \, M_\odot$ (which result in oxygen ejecta masses up to $M_\mathrm{O,ej} < 3.1 \, M_\odot$) and to increase to greater fractions for progenitor CO core masses somewhat beyond \citep{Fang2023}. This trend is consistent with the observationally inferred reference $X_\mathrm{O,ej}$ of the SN2011bm, which we estimated to have a massive CO core of $M_\mathrm{CO} > 8 \, M_\odot$: its  $5 \leq M_\mathrm{O, ej}/M_\odot \leq 10$  and $7 \leq M_\mathrm{ej}/M_\odot \leq 17$ \citep{Valenti2012} imply -- from Eq.~(\ref{eq:xoj}) -- a fraction of $ X_\mathrm{O,ej} \simeq 0.6 - 0.7$. For a fixed $M_\mathrm{O, ej}$, the greater is $X_\mathrm{O, ej}$, the lower is $M_\mathrm{ej}$. When assuming that $X_\mathrm{O,ej}= 0.7$ places a lower bound on $M_\mathrm{ej}$ for $M_\mathrm{O,ej} = 6.23 \, M_\odot$ of the most oxygen-rich Type~Ic SN explosions, then $M_\mathrm{CO} > 8.9 \, M_\odot$ follows from Eq.~(\ref{eq:mco-mej}).  

 This estimate is compatible with the window $8.4 < M_\mathrm{CO}/M_\odot < 12.4$, over which our CCSN recipe guarantees successful SNe from single star and binary-stripped star progenitors. While the rapid \citetalias{Fryer2012} and \citetalias{PS20} are also compatible with this estimate, it challenges the \citetalias{Mandel2020} and \citetalias{Mapelli2020} recipes. To achieve compatibility with this progenitor $M_\mathrm{CO}$ estimate, $M_\mathrm{CO}^\mathrm{crit}$ in \citetalias{Mapelli2020} and $M_4^*$ in \citetalias{Mandel2020} need to be lifted accordingly. 

\section{Discussion}
\label{sec:disc}

In what follows, we discuss the $M_\mathrm{CO}$ based parametrization of explodability developed in Sect.~\ref{sec:met} in the context of theoretical work.

The non-monotonicity in the final fate dependence on $M_\mathrm{CO}$ has been linked to the onset of carbon and neon burning becoming neutrino-dominated, which in turn are primarily set by $M_\mathrm{CO}$ and $X_C$ \citep{Brown2001, Sukhbold2014, ChieffiLimongi2020, Schneider21, Schneider2023, Laplace2025}. When carbon- and neon-burning become neutrino-dominated, more thermal energy leaks out of the core, which transitions from convective to radiative burning and the number as well as the size of carbon burning shells changes \citep{Sukhbold2014}. The transition from convective to radiative burning correlates with an increase in $\xi_\mathrm{2.5}$ \citep{Sukhbold2014}. However recent work by \cite{Laplace2025} made the case that the transition is not the cause for the changes in the explodability patterns and identified the mechanisms explaining the formation of the peaks in $\xi_\mathrm{2.5}$, as summarized below. When the temperature and density conditions (set by $M_{\rm{CO}}$) and the amount of nuclear fuel (which, in the case of carbon burning, is given by $X_{C}$ at carbon ignition) are such that the central burning source is strongly neutrino-dominated, the core contraction increases, leading to a large fuel-free core and, ultimately, to an increase in $M_\mathrm{Fe}$ and in $\xi_\mathrm{2.5}$. However, for even more neutrino-dominated burning at higher core masses and lower initial fuel abundance, the next nuclear burning episode ignites early, countering the core contraction and leading to a drop in $M_\mathrm{Fe}$ and $\xi_\mathrm{2.5}$. 
In what follows, we discuss how these findings relate to the threshold values $M_\mathrm{CO}^{(1)}$, $M_\mathrm{CO}^{(2)}$, and $M_\mathrm{CO}^{(3)}$ of our CCSN recipe:
\begin{itemize}
\item At low $M_\mathrm{CO}<M_\mathrm{CO}^{(1)}$, the core temperature, $T_c$, is comparatively low and $X_C$ at the end of CHeB comparatively high. Under such conditions, energy losses into neutrino cooling are lower than the energy release from core carbon burning and the core carbon burning phase is either radiation-dominated or weakly neutrino-dominated. The convective carbon burning leads to an expanded core and, due to the large amount of fuel $X_C$, the burning front does not move far outward in the mass coordinate. Ultimately, this results in a lower core density and low iron core mass at the onset of collapse. The explodability is therefore high. 
\item At $M_\mathrm{CO}^{(2)} \geq M_\mathrm{CO}\geq M_\mathrm{CO}^{(1)}$, $T_C$ is higher, therefore less fuel $X_C$ is available and the neutrino losses are greater. These lead to a neutrino-dominated core carbon burning phase. The core cools and turns radiative. The decreasing amount of fuel and the neutrino cooling accelerate core contraction and the outward progression of the burning front. The carbon burning front moves further out in the mass coordinate but stays below the effective Chandrasekhar mass. With partial degeneracy support, the core burns almost all of the $X_C$ fuel in the convective regions before ignition of radiation-dominated neon burning. After core neon-burning, the burning front quickly burns the former convective region, moving out far in the mass coordinate. This leads to the growth of a large and dense fuel-free core. The explodability is therefore low.  

\item At greater $M_\mathrm{CO}>M_\mathrm{CO}^{(2)}$, core carbon burning is even more neutrino-dominated,  due to a high $T_c$ and low $X_C$. The burning phase proceeds faster, and the core contraction is even more accelerated. The burning front moves further out in the mass coordinate, until it exceeds the effective Chandrasekhar mass. The contraction leads to an early core neon ignition, before all the carbon in the core is burnt. This next (radiation-dominated) core-burning stage suppresses nuclear burning of the carbon-burning front above, preventing it from moving far outward. Ultimately, this results in a low-mass iron core and high explodability.

\item At high $M_\mathrm{CO} \geq M_\mathrm{CO}^{(3)}$, it is the core neon burning phase that becomes neutrino-dominated. The neutrino cooling leads to a quickly contracting radiative core and to an accelerated progression of the burning front above. The burning front again moves further out but stays below the effective Chandrasekhar mass. With partial degeneracy support, the core burns most of the neon fuel before the ignition of radiation-dominated oxygen burning. The central burning of the large oxygen core leads to an enhanced growth of the silicon-rich core, as the burning front moves out in the mass coordinate. These lead to a massive iron core and low explodability.  
\end{itemize}

\section{Conclusion}
\label{sect:con}

The outcome of a core-collapse supernova (CCSN) is a complex multidimensional phenomenon, which is appropriately addressed using computationally expensive 3D CCSN simulations. However, population synthesis and many other astrophysical studies require efficient models for predicting CCSN outcomes at scale. In this work, we have formulated explodability criteria for the neutrino-driven SN mechanism that allow us to predict the final fate (successful or failed SN) already at the pre-SN stage. Then, we used stellar evolution models of single stars and binary-stripped stars to construct a CCSN recipe that is based on the carbon-oxygen core mass, $M_\mathrm{CO}$, and metallicity, $Z$.

To obtain the criteria, we parametrized the explodability by compiling a heterogeneous set of $\simeq 3900$ single star, binary-stripped, and accretor star pre-SN models and identifying upper and lower thresholds  in pre-SN stellar structure variables that coincide with failed and successful SNe, respectively, as predicted by the semi-analytical \citetalias{M16} SN model. The explodability criteria evaluate the SN progenitor using multiple diagnostic scalar variables: the compactness, $\xi_\mathrm{2.5}$; the central specific entropy, $s_c$; $M_\mathrm{CO}$;  the $\mu_4 M_4$ variable, which relates to the accretion luminosity above the proto-neutron star (PNS); and the $\mu_4$ variable, which relates to the mass accretion rate above the PNS. These probe the SN progenitor profile structure at four different mass coordinates. 
Our explodability scheme achieves a predictive accuracy of >99\% agreement with final fate predictions by \citetalias{M16}.

A successful SN leaves behind either a neutron star (NS) or a fallback black hole (BH) remnant. We find that fallback BH formation, as predicted by \citetalias{M16}, can be excluded, when $\xi_\mathrm{2.5}$ is either critically low, low compared to $\mu_4 M_4$, or when $M_4 > 0.6 \,  M_\mathrm{CO}$. Fallback BH formation occurs at a frequency of $\sim$~0.15 over our exploding models; namely, a NS is the multiple times more likely compact remnant type. The fallback BH formation model is a particularly uncertain part of our predictive framework, since it has neither been validated against 3D CCSN simulation outcomes, nor against observations. Our explodability formalism likely underestimates BH formation in  successful SNe, as it does not take into account two out of the three formation channels (see Sect.~\ref{sec:fallback}). Accounting for these requires further development of the framework modeling explosive BH formation in \citetalias{M16} prior to the statistical model-building pursued in this work.

We compared our explodability scheme as well as commonly used alternative SN models with the outcomes of 3D CCSN simulations from the archives of the Garching and the Monash groups. Our pre-SN explodability criteria achieve an agreement of 86\% over 29 simulation outcomes. The cases of disagreement are with stellar models for Population~III stars and a rapidly rotating long gamma-ray burst progenitor that are not covered by our compilation of stellar progenitor models. 
Our scheme is more optimistic about successful CCSN explosions by the neutrino-driven mechanism than both the criterion introduced in \cite{Ertl2016}  -- which is based on a separation line in the $(\mu_4 M_4, \mu_4)$ plane -- and the compactness criterion, $\xi_\mathrm{2.5} > 0.45$ \citep{OConnor2021}. With \citetalias{M16}, we find that failed and successful SNe coexist over a wide range in $\xi_\mathrm{2.5}$. Since a separation line in the $(\mu_4M_4, \mu_4)$ plane that would segregate the exploding from the non-exploding models cannot be drawn, the criterion introduced in \cite{Ertl2016} cannot explain the distribution of CCSN outcomes predicted by \citetalias{M16} among our sample of SN progenitor models; nor can it explain the distribution of 3D CCSN simulation outcomes considered in this work.

Using the explodability criteria and a subset of single star and binary-stripped star models adopting the same input physics, we constructed a CCSN recipe that predicts the compact remnant type (direct collapse BH, fallback BH or NS) based on $M_\mathrm{CO}$, $Z$ and mass transfer history class (single star, Case~A, Case~B, or Case~C donor) already at the end of core-helium burning. This is possible, because all pre-SN stellar structure variables relevant for evaluating our explidability criteria show bimodal trends -- characterized by two peaks and an intervening valley -- as a function of $M_\mathrm{CO}$ of single stars and binary-stripped stars within the wide range  of $6.1 < M_\mathrm{CO}/M_\odot < 15.4$, over which we found successful and failed SNe to coexist.  To obtain the CCSN recipe, we mapped out the boundaries in $M_\mathrm{CO}$ at which the expected final fate transitions from a successful SN to a failed SN and vice versa. In the case of a successful SN, the compact remnant type (NS or fallback BH) is predicted probabilistically. The recipe is made publicly available and can be readily implemented for binary population synthesis and other studies. 

We find that the failed SN windows in $M_\mathrm{CO}$ of our single star and binary-stripped star models shift toward larger values when $Z$ increases or binary mass transfer (MT) sets in earlier. Envelope mass loss by stellar winds (which are enhanced by a greater $Z$) or by stable MT to a companion star result in lower-mass helium cores. Lower-mass helium cores are cooler in the center, which during core-helium burning leads to a later ignition of the strongly temperature-dependent $^{12}\mathrm{C}\left(\alpha, \gamma\right) ^{16}\mathrm{O}$ reaction that uses up carbon. This leaves behind more carbon at core-helium exhaustion. With a greater central carbon mass fraction, $X_C$, at core-helium exhaustion, $M_\mathrm{CO}$ needs to be greater for having the transition from a radiation-dominated to a neutrino-dominated core-carbon burning phase, which anticipates pre-SN core structures that likely result in failed SNe.

Comparing 3D CCSN simulation outcomes with our and other $M_\mathrm{CO}$-based CCSN recipes over progenitors with $M_\mathrm{CO} < 16 \, M_\odot$, the number of disagreements is:\ one for \citetalias{PS20},  two for \citetalias{Mandel2020} and ours, three for \citetalias{Fryer2012}, and five for \citetalias{Mapelli2020},  respectively. The distinctive feature of our CCSN recipe compared to alternative models is that it guarantees explosions for $8.4 \leq  M_\mathrm{CO}/M_\odot \leq 12.4$, independent of $Z$ and binarity, with a more likely NS than fallback BH remnant. Moreover, it admits CCSN explosions of progenitors with $M_\mathrm{CO}$ as high as 15.4 $M_\odot$.

We tested our and other $M_\mathrm{CO}$-based CCSN recipes against observations that constrain the $M_\mathrm{CO}$ of Type~IIP, Type~IIn, Type~IIb and Ib, and Type~Ic SN progenitors. We arrived at the following main conclusions.

Since the pre-SN luminosity is a direct tracer of $M_\mathrm{CO}$, we inferred that the missing red supergiant (RSG) problem \citep{Smartt2009, Smarttetal2009} is manifested over a CO core mass range of at least $6 \leq M_\mathrm{CO}/M_\odot \leq 7.8$.  Our recipe is consistent with the most luminous Type~IIP SN progenitors and partially addresses the missing RSG problem by failed SNe. The \citetalias{Fryer2012} model addresses the missing RSG problem over a similar range ($6 \leq M_\mathrm{CO}/M_\odot \leq 7$) as our model. The \citetalias{PS20} model is consistent with the most luminous Type~IIP SN progenitors and can address the missing RSG problem by failed SNe entirely, provided that the RSG progenitors have a low enough $X_C$ at the end of core-helium burning. 

With the Type~Ic SN2011bm \citep{Valenti2012}, the most oxygen-rich Type~Ic SNe that have been revealed by nebular line spectroscopy \citep{Fang2022} and the Type~IIn SN2010jl \citep{Smith2010} -- when assuming a single star progenitor for this transient --, we find putative evidence for explosions of stars with $M_\mathrm{CO} > 8.2 \, M_\odot$, which is in agreement with  predictions of our model. Such explosions are incompatible with the default versions of the \citetalias{Mapelli2020} and \citetalias{Mandel2020} SN models.
The \citetalias{Mapelli2020} SN model is not equipped to address the missing RSG problem by failed SNe and simultaneously remain consistent with the lower bound on the Type~Ic SN2011bm ejecta mass as well as with the most oxygen-rich Type~Ic SNe. 
In the \citetalias{Mandel2020} SN model, the threshold value for guaranteed BH formation by direct collapse needs to be lifted to $ > 9 \, M_\odot$  to remain consistent with the aforementioned transients. \citetalias{Mandel2020} cannot address the missing RSG problem by failed SNe.
The most luminous progenitors of Type~IIb and Ib SNe allow for a  $M_\mathrm{CO} > 6 M_\odot$ within the observational uncertainty bounds,  thereby challenging the \citetalias{Fryer2012} model, which predicts failed SNe for $6 \leq M_\mathrm{CO}/M_\odot \leq 7$ regardless of binarity.
Furthermore, the compact remnant mass prescription of \citetalias{Fryer2012} is incompatible with the inferred ejecta mass of the Type~Ic SN2011bm. 

As a bottom line, given that the $M_\mathrm{CO}$ estimates of observed stripped-envelope SNe involve wide error bars, that inferences based on the nebular spectroscopy of Type~Ic SN remnants rest on assumptions and that the progenitors of Type~IIn SNe are not well constrained at present, we hesitate to overinterpret these comparisons. Nevertheless, we aim to convey the idea of benchmarking CCSN recipes using observed transients. Since binary-stripped stars are the expected progenitors of both components of binary BH mergers, we expect that our CCSN recipe will result in a suppression of the predicted binary BH merger rates, compared to previous estimates. 

\section*{Data availability}
The code for predicting the final fates of stars and discriminating their compact remnant types based on the pre-SN properties using the explodability and fallback BH formation criteria introduced in this work is available at \url{https://zenodo.org/records/15350140} in form of a Jupyter Notebook written in Python. Under the same link but in another Jupyter Notebook written in Python is the code implementing the $M_\mathrm{CO}$--based CCSN recipe for binary population synthesis. The CCSN recipe is also available written in C++ through the \textsc{compas} (\url{https://compas.science/}) and the \textsc{sevn} (\url{https://sevncodes.gitlab.io/sevn/index.html}) population synthesis codes.

\begin{acknowledgements}
The data on the progenitors of the 3D CCSN simulations performed in the Garching group and on the observed Type~IIn SN progenitors have kindly been supplied by Daniel Kresse and by Nancy Elias-Rosas, respectively.
KM thanks Philipp Podsiadlowski, Hans-Thomas Janka, Daniel Kresse, Keiichi Maeda, Takashi Nagao, Nancy Elias-Rosas, Giuliano Iorio, Erika Korb, Michela Mapelli, Evan O'Connor, Bob Fisher, Ósmar Rodríguez, Koh Takahashi and Raphael Hirschi for helpful discussions.
We thank the anonymous referee for their helpful comments. KM, FRNS, FKR, and EL acknowledge support by the Klaus Tschira Foundation. This work has received funding from the European Research Council (ERC) under the European Union’s Horizon 2020 research and innovation programme (Grant agreement No.\ 945806), and is supported by the Deutsche Forschungsgemeinschaft (DFG, German Research Foundation) under Germany’s Excellence Strategy EXC 2181/1-390900948 (the Heidelberg STRUCTURES Excellence Cluster).  IM acknowledges support from the Australian Research Council (ARC) Centre of Excellence for Gravitational Wave Discovery (OzGrav), through project number CE230100016. EL acknowledges support through a start-up grant from the Internal Funds KU Leuven (STG/24/073) and through a Veni grant (VI.Veni.232.205) from the Netherlands Organization for Scientific Research (NWO). BM acknowledges support by the ARC through Discovery
Project (DP) DP240101786.
AH acknowledges support by the ARC through DPs
DP240101786, DP240103174, though ARC LIEF project LE230100063, and by
the Alexander von Humboldt Foundation through a Research Award.
\end{acknowledgements}
  

\bibliographystyle{aa}
\bibliography{biblio}

\begin{appendix}

\section{Supplementary materials}

\subsection{Progenitors of 3D CCSN simulations}

Tables \ref{Tab:garching} and \ref{Tab:monash} reference the 3D CCSN simulations from the Garching and Monash groups considered in this work. 

\begin{table*}[h!]
  \begin{center}
  \caption{Properties of the stellar progenitor models in the 3D CCSN simulations performed in the Garching group. The final fate (ff) column indicates the outcome of the simulation (0: failed SN, 1: successful SN).}
    \begin{tabular}{l l l l l l l l l}       
    \hline\hline  
      
       model & ff & $M_\mathrm{CO}/M_\odot$ & $\xi_\mathrm{2.5}$ & $\frac{s_c}{(k_B \, N_A)}$ & $\mu_4 M_4$ & $\mu_4$ & Ref. progenitor & Ref. 3D CCSN sim.\\
      \hline
      s9.0 & 1 & 1.40 & $4 {\cdot} 10^{-5}$ & 0.53 & $2 {\cdot} 10^{-5}$ & $2 {\cdot} 10^{-5}$ & \cite{Woosley2015} & \cite{Melson2020} \\
      z9.6 & 1 & 1.37 & $8 {\cdot} 10^{-5}$ & 0.89 & $3 {\cdot} 10^{-5}$ & $2 {\cdot} 10^{-5}$ & Heger (2012), priv. comm. & \cite{Melson2015}  \\
      s12.28 & 1& 2.23 & 0.0312 & 0.61 & 0.061 & 0.039 & Yadav (2023), priv. comm. 
      & \cite{Janka2024}\\
      m15 & 1 & 3.24 & 0.106 & 0.68 & 0.073 & 0.046 & \cite{Heger2005} &  \cite{Summa2018} \\
      s18.88 & 1 & 4.45 & 0.283 & 0.87 & 0.200 & 0.110 & \cite{Yadav2020} & \cite{Bollig2021} \\
      s20 & 1 & 4.98 & 0.285 & 0.86 & 0.201 & 0.110 & \cite{Woosley2007}  & \cite{Melson2015b}\\
      \hline
      s40 & 0 & 13.59 & 0.544 & 1.05 & 0.470 & 0.207 & \cite{Woosley2007} & \cite{Walk2020}  \\
      u75 & 0 & 31.16 & 0.882 & 1.53 & 0.793 & 0.321 & \cite{Woosley2002} & \cite{Walk2020} \\      
      \hline
    \end{tabular}
  \label{Tab:garching}
  \end{center}  
\end{table*}

\begin{table*}[h!]
  \begin{center}
  \caption{Properties of the stellar progenitor models in the 3D CCSN simulations performed in the Monash group. The final fate (ff) column indicates the outcome of the simulations (0: failed SN, 1: successful SN). 
  }
    \begin{tabular}{l l l l l l l l l}       
    \hline\hline  
      
       model & ff & $M_\mathrm{CO}/M_\odot$ & $\xi_\mathrm{2.5}$ & $\frac{s_c}{(k_B \, N_A)}$ & $\mu_4 M_4$ & $\mu_4$ & Ref. progenitor & Ref. 3D CCSN sim. \\
      \hline
      s9.5 & 1 & 1.49 & $1.6 {\cdot} 10^{-5}$ &  0.53 & 0.006 & 0.004 & \cite{M16} & Sykes et al., in prep.\\
      z9.6 & 1 & 1.37 & $7.6 {\cdot}  10^{-4}$ &        0.89 &  $6.4 {\cdot} 10^{-5}$ &      $4.7 {\cdot} 10^{-5}$ & \cite{Heger2010} & \cite{Mueller2016} \\
      s11.5 & 1 & 1.48 &        $1.2 {\cdot} 10^{-5}$ & 0.51 &  0.004 & 0.003 & \cite{M16} & Sykes et al., in prep. \\
      s11.8 & 1 & 1.61 &        $5.9 {\cdot} 10^{-5}$ & 0.53 &  0.022 & 0.016 &        \cite{Banerjee2016} & \cite{Mueller2019} \\
      z12   & 1 & 1.79 & 0.011    & 0.52  & 0.030 & 0.019 & \cite{Heger2010} & \cite{Mueller2019} \\
      s12.5 & 1 & 2.05 & 0.020   & 0.70 & 0.140 & 0.087 & \cite{M16} & \cite{Mueller2019} \\
      s14.07 & 1 &  2.56 & 0.130 & 0.73 & 0.263 & 0.163 & \cite{M16} & unpublished \\
      s15s7b2 & 1 & 2.48 & 0.088 & 0.73 & 0.216 & 0.150 & \cite{Woosley1995} & \cite{Powell2024} \\
      m15b2 & 1 & 2.91 &        0.087 & 0.62 &  0.128 & 0.087 & \cite{Heger2005} & \cite{MV2020} \\
      s16.9 & 1 & 4.14 &        0.144 & 0.89 &  0.175 & 0.104 & \cite{Schneider2019} & \cite{Varma2023} \\
      s18 & 1 &  3.83  &        0.244 & 0.83 &  0.353 & 0.201 & \cite{M16} & \cite{Mueller2017} \\
      s24 & 1 &  6.08 & 0.261 & 0.85 & 0.333 & 0.182 & \cite{M16} & Sykes et al., in prep. \\
      z40 & 1 & 12.92 & 0.638 & 1.15 & 1.516 & 0.698 & \cite{Heger2010} & \cite{Chan2020} \\
      z85 & 1 & 31.25 & 0.856 & 1.57 & 1.673 & 0.691 & \cite{Heger2010} & \cite{Powell2021} \\
      he2.8 & 1& 1.46 & --    & 0.89     & --     & -- & \cite{MuellerG2018} & \cite{Mueller2019mnras} \\ 
      he3 & 1 & 1.91 & 0.016 & 0.64 & 0.118 & 0.078     & \cite{M16} & \cite{Mueller2019mnras} \\
      he3.5 & 1 & 1.81 & -- & 0.68 & 0.012 & 0.007 & \cite{MuellerG2018} & \cite{Mueller2019mnras} \\
      y20 & 1 & 8.16 & 0.223 & 1.32 & 0.140 & 0.082 & \cite{Yoon2017b} & \cite{Powell2020} \\
      m39 & 1 & 20.95 & 0.364 & 1.06 & 0.443 & 0.225 & \cite{AguileraDena2018} & \cite{Powell2020} \\
      \hline
      s14 &  0 & 2.53 & 0.112  & 0.72 & 0.235 & 0.147 & \cite{M16} & Sykes et al., in prep. \\
      z100 & 0 & 42.44 &        0.400 & 0.93 &  0.671 & 0.331 & \cite{Heger2010} & \cite{Powell2021} \\   
      \hline
    \end{tabular}
  \label{Tab:monash}
  \end{center}  
\end{table*}

Unless otherwise specified, the SN progenitors are non--rotating. The model naming acronyms encode metallicity (``s'' for solar, ``z'' for Population~III), ZAMS mass (e.g., 9.5 for $M_\mathrm{ZAMS} = 9.5 \, M_\odot$) and binary-stripped stars ('he' for a helium star at $Z = Z_\odot$). Models starting with ``m'' are rotating, those with ``y'' are stripped star models.

We add remarks on the simulations performed in the Garching group:
For the models s12.28 and s18.88, the final minutes of convective
oxygen-shell burning before the core collapse have been simulated in 3D,
yielding large-scale and large-amplitude progenitor perturbations \citep{Yadav2020}. When starting from 1D initial conditions, these two models did not explode. The explosion of the model m15 was aided by rapid
progenitor rotation \citep{Summa2018} and the explosion of the s20 progenitor by a slight modification of the neutral-current neutrino-nucleon scattering opacities \citep{Melson2015}.

We also remark on the simulations performed in the Monash group:
For the low--mass ($M_\mathrm{CO} < 2 \, M_\odot$) progenitors he2.8, he3.5, s9.5, z9.6, s11.5 and s11.8 and the high--mass ($M_\mathrm{CO} > 8 \, M_\odot$) progenitors y20, z40, z85 and z100 and m39, CCSN simulations were performed over spherically symmetric stratifications obtained from 1D progenitor models and did not include magnetic fields. With the exception of z100, all of these exploded. In the intermediate $M_\mathrm{CO}$--range, perturbations were introduced into the originally spherically symmetric stratifications of the progenitors he3, z12, s12.5 and s14.07. These models exploded in 3D simulations without the enhancing effect of magnetic fields. 3D CCSN simulations were carried out with magnetic fields over the spherically symmetric stratifications of s14, s15s7b2 and of the very slowly rotating s16.9 model. 

The effect of perturbations and of magnetic fields was studied systematically upon the SN progenitors s18 and m15b2 (slowly rotating). These did not explode when starting from spherically symmetric progenitor stratifications and not including magnetic fields, but exploded once magnetic fields were introduced. The 3D CCSN simulation without magnetic fields but over a perturbed s18 profile structure statification also resulted in an explosion. 

\subsection{The $2.5 \, M_\odot$ mass coordinate and $M_\mathrm{Fe}$}

\begin{figure}
  \centering
  \includegraphics[width=0.45\textwidth]{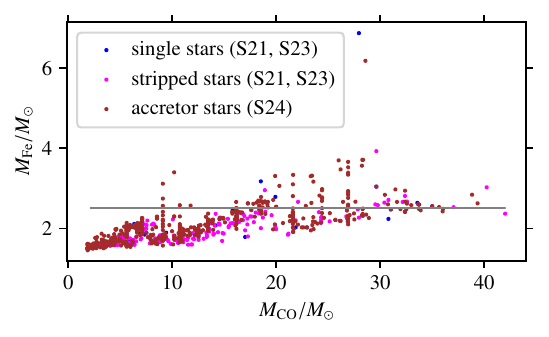}
  \caption{Trends of iron core mass $M_\mathrm{Fe}$ with carbon-oxygen core mass $M_\mathrm{CO}$ for single, binary-stripped and accretor star pre-SN models from \cite{Schneider21, Schneider2023, Schneider2024}. The $M = 2.5 \, M_\odot$ mass coordinate that is typically used for $\xi_\mathrm{M}$ is marked by a grey solid line.}
\label{fig:comp-mfe}
\end{figure}

Fig.~\ref{fig:comp-mfe} shows the trends of iron core mass $M_\mathrm{Fe}$ with carbon-oxygen core mass $M_\mathrm{CO}$, and tests how frequently the iron core is more massive than $M = 2.5 \, M_\odot$, which is the mass coordinate that is often chosen to evaluate the compactness parameter $\xi_M$.

\subsection{Fits of explodability proxies}
\label{app:explod_prox}
\begin{figure*}
      \centering
      \includegraphics[width=0.9\textwidth]{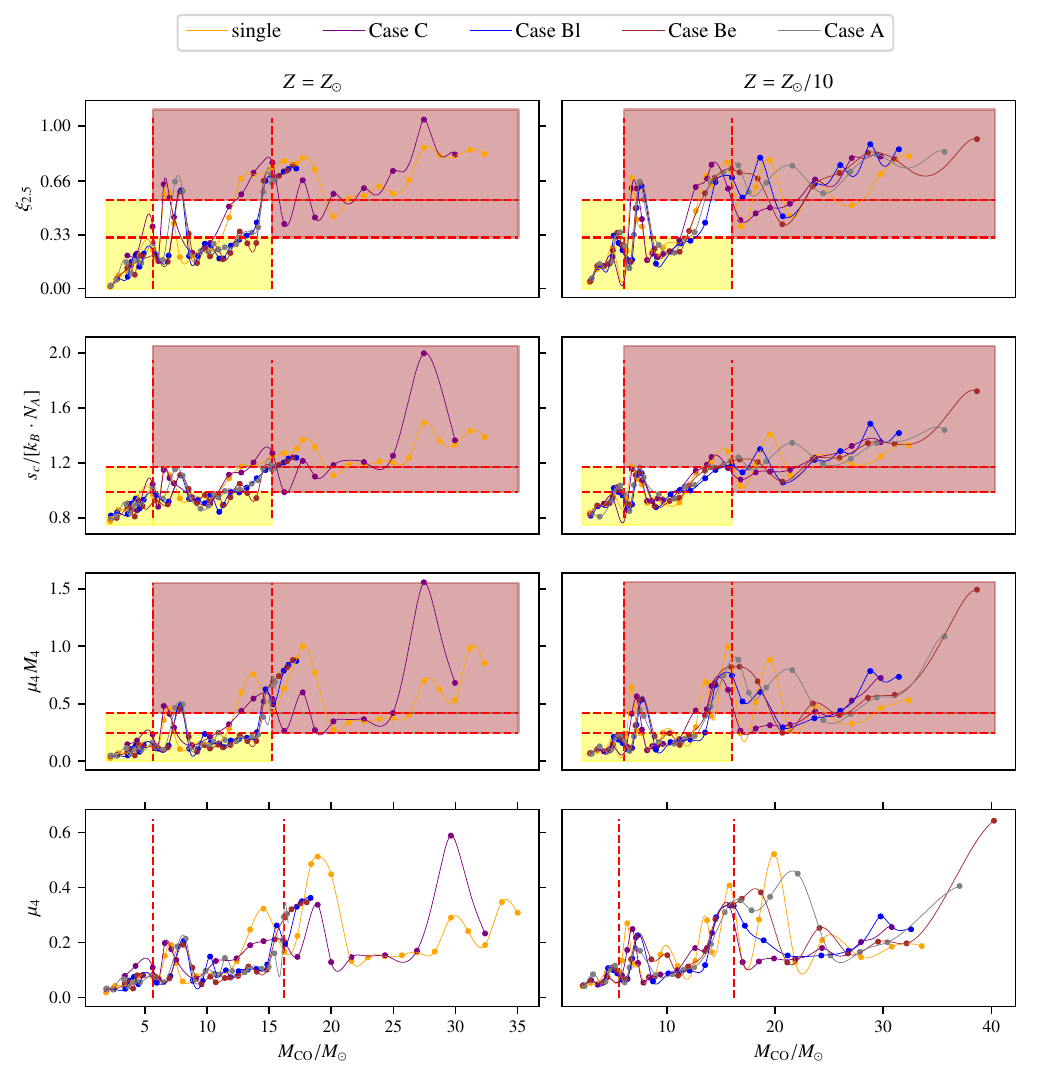}
      \caption{Dependence of the explodability parameters 
      $\xi_\mathrm{2.5}$, $s_c$, $\mu_4 M_4$ and $\mu_4$ of single and binary-stripped stars (Case~A, Case~Be, Case~Bl and Case~C donors) on $M_\mathrm{CO}$ at $Z=Z_\odot$ and at $Z=Z_\odot /10$. To obtain these, the data points (circles) from \citetalias{Schneider21} and \citetalias{Schneider2023} have been fitted using GPR models (solid lines). 
      The red vertical lines indicate the lower (upper) threshold in $M_\mathrm{CO}$ 
      below (beyond) which only successful (failed) SNe occur, while the red horizontal lines indicate the lower and upper thresholds in $\xi_\mathrm{2.5}$, $\mu_4M_4$ and $s_c$, respectively. The background colors show the resulting final fate classification based on a single explodability parameter: successful SNe (yellow), failed SNe (brown) and unclassified (blank).}  
    \label{fig:proxies-fit}
\end{figure*}

Fig.~\ref{fig:proxies-fit} shows the fits of explodability proxies using GPR models. Since the pioneering work by \cite{Sacks1989}, GPR  has been a standard method for emulation tasks because of its flexibility, smoothness and the regulatory effect of the Gaussianity assumption. 
The supervised learning method is used to train GPR models to predict the outputs $y_1, ..., y_n$ given the inputs $x_1, ..., x_n$. If $n$ is the size of the training data set, then GPR interprets the output data
as a random sample drawn from an $n$-dimensional multivariate normal 
\begin{equation}
    \mathcal{N}_n (\mathbf{\mu}, \mathbf{\Sigma}) = \frac{\exp \left(-\frac{1}{2}(\mathbf{X} - \mathbf{\mu}) ^T \mathbf{\Sigma}^{-1}\left( \mathbf{X} - \mathbf{\mu} \right)\right)}{\sqrt{(2 \pi)^n | \mathbf{\Sigma} |}} 
\end{equation}
which has the mean vector $\mathbf{\mu}$ and the functional form of the $n \times n$ covariance matrix $\mathbf{\Sigma}_\mathrm{i,j} = \mathrm{Cov}[x_i, x_j]$ as free parameters that need to be set before training. While $\mathbf{\mu}$ is typically set to the zero vector, different kernel functions $k(x_i, x_j)$ are available\footnote{For a selection of kernel models, see, e.g., \url{https://scikit-learn.org/stable/modules/generated/sklearn.gaussian_process.GaussianProcessRegressor.html}} 
that specify $\mathbf{\Sigma}$. For example, the Matérn kernel takes the following form:
\begin{equation}
    k(x_i, x_j) = \frac{1}{\Gamma(\nu) \, 2^{\nu -1 }} \left[ \frac{\sqrt{2 \nu}}{l} \, d \left(x_i, x_j \right) \right]^\nu K_\nu \left( \frac{\sqrt{2 \nu}}{l} \, d \left(x_i, x_j \right) \right),
\end{equation}
where $d(x_i, x_j)$ is the Euclidean distance between $x_i$ and $x_j$, $K_\nu$ is the modified Bessel function, and $\Gamma$ is the Gamma function. 
For more technical detail, refer, for instance, to \cite{Rasmussen2004} and for a visual exploration, to \cite{goertler2019a}.

\subsection{Deterministic model for fallback BH formation}
\label{app:det_fb}

\begin{figure}
  \centering
  \includegraphics[width=0.45\textwidth]{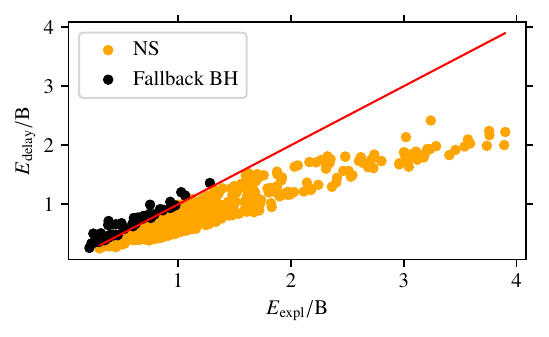}
  \caption{Discrimination of the remnant type (NS or fallback BH) in a successful CCSN explosion using the $E_\mathrm{delay} > E_\mathrm{expl}$ condition for fallback BH formation. This criterion holds exactly over the \citetalias{Schneider21}, \citetalias{Schneider2023}, \citetalias{Schneider2024} and \citetalias{Temaj} SN progenitors (evaluated in the panel), and only approximately over H16.}
\label{fig:fallback}
\end{figure}

\begin{figure*}
      \centering
      \includegraphics[width=0.45\textwidth]{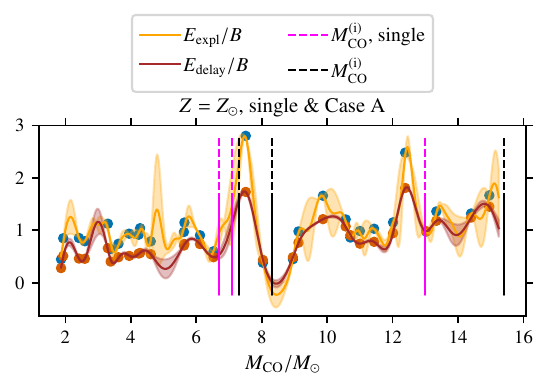}
      \includegraphics[width=0.45\textwidth]{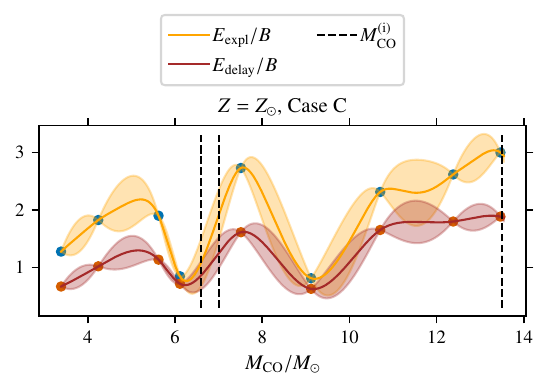}
      \includegraphics[width=0.45\textwidth]{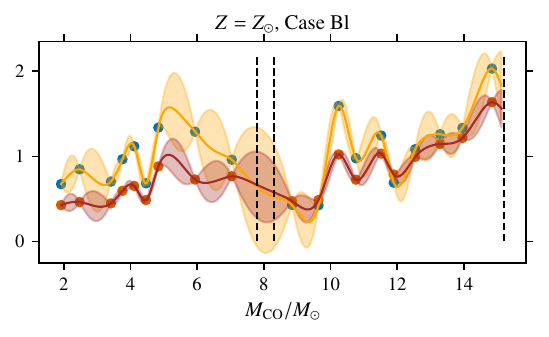}
      \includegraphics[width=0.45\textwidth]{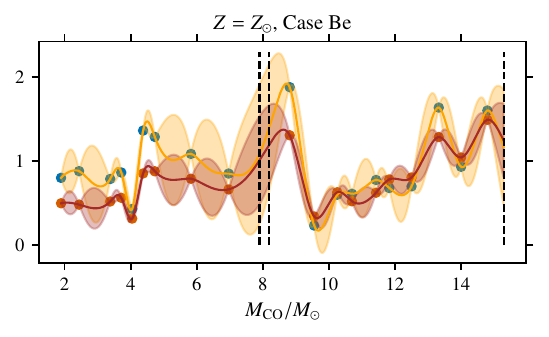}
      \caption{BH formation by fallback in SN explosions of single and binary-stripped stars at $Z = Z_\odot$, as predicted by the deterministic criterion $E_\mathrm{delay} - E_\mathrm{expl} > 0$. The dots show the $E_\mathrm{expl}$ and $E_\mathrm{delay}$ predictions with \citetalias{M16}, which is used to explode the SN progenitor models from the \citetalias{Schneider21} sample, as a function of progenitor $M_\mathrm{CO}$. These are fitted      
       using GPR regressors for each of the SN progenitor types as classified by MT history (single, Case~A, Case~Be, Case~Bl, Case~C). Non-exploding single and binary-stripped star models were removed from the \citetalias{Schneider21}  sample. The direct BH formation boundaries predicted by our CCSN recipe are indicated by vertical dashed lines.  Due to the sparse sampling of the \citetalias{Schneider21} stellar models, the prediction intervals (shaded regions) are wide and overlap, not allowing for a faithful prediction of the remnant type using our  deterministic fallback BH formation criterion.} 
    \label{fig:fallback-mco}
\end{figure*}

The condition for BH formation from \cite{M16} applies during the explosion phase: if the diagnostic energy $E_\mathrm{diag} < 0$, then the gravitational binding energy of the matter enclosed by the mass shell up to which the revived shock has expanded is greater than the kinetic energy of the explosion. In this case, a BH forms by fallback of  matter onto the PNS. 

With \citetalias{M16}, we find that the remnant type can be predicted by comparing two characteristic energies during the explosion phase: the final explosion energy $E_\mathrm{expl}$ and the energy term $E_\mathrm{delay}$ which -- within the limitations of a 1D formulation -- by construction accounts for the coexistence of outflows and downflows in the region surrounding the PNS during the explosion phase.  $E_\mathrm{delay}$ is an auxiliary variable defined implicitly from Eq.~(42) and (43) in \cite{M16}, which are used to calculate the evolution of $E_\mathrm{diag}$ as the revived shock moves outward in mass shell. With the fallback BH formation criterion 
\begin{equation}
\label{eq:fb_eqn}
    E_\mathrm{delay} > E_\mathrm{expl},
\end{equation}
an accuracy of 100\% for the remnant type prediction (NS versus fallback BH) is achieved over the \citetalias{Schneider21}, \citetalias{Schneider2023}, \citetalias{Schneider2024} and \citetalias{Temaj} stellar models (shown in Fig. \ref{fig:fallback}), and of 93\%  over the H16 models. This criterion is sensitive to our  parameter choice for the \citetalias{M16} SN code.

This empirical criterion is made plausible by the following reasoning. Physically, one expects fallback to be determined primarily by the ratio of initial explosion energy $E_\mathrm{ini}$ at shock revival to the envelope binding energy $E_\mathrm{bind}$, with a drastic increase of fallback once $E_\mathrm{bind}$ gets close to $E_\mathrm{diag}$. $E_\mathrm{delay}$ is expected to correlate well with $E_\mathrm{ini}$, say $E_\mathrm{delay} = \eta \cdot E_\mathrm{ini}$, where $\eta < 1$ is a parameter measuring the amount of energy that is dissipated during shock propagation. To zeroth approximation,  the final explosion energy is given by $E_\mathrm{expl} =  E_\mathrm{ini} + E_\mathrm{delay} - E_\mathrm{bind}$. Therefore $E_\mathrm{expl} - E_\mathrm{delay} < 0$ implies that $E_\mathrm{expl}/E_\mathrm{ini} < \eta$. In other words, fallback BH formation is expected to occur if the explosion has lost a critically large fraction of its initial energy. That the occurrence of BH formation by fallback within \citetalias{M16} is tied to the auxiliary variable $E_\mathrm{delay}$ could be coincidental, since $E_\mathrm{delay}$ does not have a direct physical significance.

As shown in Fig.~\ref{fig:fallback-mco}, single and binary-stripped stars of different MT classes show complex, oscillatory patterns in the dependence of the $E_\mathrm{expl}$ and $E_\mathrm{diag}$ variables on $M_\mathrm{CO}$. In principle, windows in $M_\mathrm{CO}$ can be mapped out over which Eq.~(\ref{eq:fb_eqn}) is satisfied as a function of MT history and $Z$. The GPR regressor fit curves are consistent with the hypothesis that fallback BH formation is not randomly distributed but occurs over windows in $M_\mathrm{CO}$, and that the width and position of the windows varies with MT class. However, in contrast to H16, the sampling of the $M_\mathrm{CO}$ axes in \citetalias{Schneider21} and \citetalias{Schneider2023} is too sparse for drawing faithful conclusions. 

\subsection{Practical remarks on using the CCSN recipe}
\label{app:ccsn_rec}

\subsubsection{Mass transfer class assignment}
\paragraph{LBV stars:}
For stars undergoing the LBV phase of enhanced mass loss  over a short (thermal or faster) timescale, we suggest to classify their explodability using the routines for Case~B donors rather than those for single stars. 

\paragraph{Repeated MT episodes:}
In the case of repeated MT episodes, in particular Case~AB and Case~BC donors, we suggest to proceed as follows. For Case~AB donors, whether MT class A or B is assigned to the star, does not substantially change the final CCSN outcome, since the pre--SN properties of Case~A and B progenitors are similar. For Case~BC donors, we suggest to use the first MT episode for the classification, because Case~C progenitors are closer to single stars, and the binary interaction effects are likely more adequately accounted for by the critical values we found for Case~B donors. 

\paragraph{Partial envelope stripping:}
In the case of partial stripping of the hydrogen-rich envelope rather than its complete removal, one of the ways to proceed is to set a threshold value in the fraction of envelope mass that is removed by stable MT, above (below) which the star is classified as a stripped (single) star, for example 50\%. Another way is to linearly interpolate  the $M_\mathrm{CO}^{(i)}(Z)$--thresholds of the single and the donor stars, depending on the fraction of envelope mass removed. 

\paragraph{Accretor stars:}
The CCSN recipe is built for single stars and binary-stripped stars. For accretor stars that gain mass by Roche lobe overflow from a donor during the main sequence (MS) evolution, we assume that their explodability can be approximated using the single star routines, since their cores rejuvenate. This approximation breaks down for stars accreting mass during their post-MS evolution.

\subsubsection{Breakdown of the CCSN recipe}
Application of our CCSN recipe for $Z \notin (Z_\odot/10, Z_\odot)$ implies extrapolation. The recipe breaks down below some cutoff $Z_\mathrm{cutoff} < Z_\odot /10$, where $Z_\mathrm{cutoff}$ is presumably higher than zero (Population~III stars). At high enough super-solar metallicities, BH formation by direct collapse at high $M_\mathrm{CO}$ is expected to be precluded because of the higher effective Chandrasekhar mass required for the collapse to set in. Since the CCSN recipe does not explicitly take this physical effect at increasing $Z$ into account, it will presumably break down in the regime of high $M_\mathrm{CO}$ and high super-solar $Z$. Practically, however, this may be not a concern, since at sufficiently high super-solar $Z$, winds are strong enough to erode the CO core such that stars will not end up having such high $M_\mathrm{CO}$ cores at which the CCSN recipe would (erroneously) predict direct BH formation.

\subsection{Other $M_\mathrm{CO}$-based CCSN recipes}
\label{sec:ccsn_recipes}

In the following, we summarize the $M_\mathrm{CO}$-based CCSN recipes that are referenced in this work and compared to ours.

\subsubsection{MM20}
In \cite{Mandel2020}, the remnant mass and type are assigned using probabilistic formulae based on calibrated threshold values $M_i^*$, with $i = 1, \dots, 4$, in $M_\mathrm{CO}$. Core collapse is predicted to result in NS formation for $M_\mathrm{CO} \leq M_1^*$. BH (formed either by fallback or directly) and NS remnants coexist for $M_1^* < M_\mathrm{CO} \leq M_3^*$. For $M_3^* < M_\mathrm{CO} \leq M_4^*$, BH formation, either directly or by fallback, is guaranteed. For $M_\mathrm{CO} > M_4^*$, only direct BHs form. The default threshold values are $M_1^* = 2 \, M_\odot$, $M_3^* = 7 \, M_\odot$ and $M_4^* = 8 \, M_\odot$. \citetalias{Mandel2020} is used as CCSN recipe in \textsc{compas}, for example.

While both \citetalias{Mandel2020} and our CCSN recipe are constructed based on outcomes of the \citetalias{M16} SN code, there are two main differences. The first concerns the parameter choice for \citetalias{M16}: \citetalias{Mandel2020} use the default settings from \cite{M16}, except for a higher accretion efficiency ($\zeta = 0.8$) and a different calculation of the final mass cut upon BH formation by fallback in a successful SN explosion. The second difference concerns the SN progenitor models: \citetalias{Mandel2020} is based on the H16 single star models at $Z=Z_\odot$, and uses a randomized scheme that follows gross trends of the compact remnant masses $M_\mathrm{rem}$, predicted by \citetalias{M16}, with $M_\mathrm{CO}$ of the H16 SN progenitors. 

\subsubsection{M20} 
According to \cite{Mapelli2020}, there is no coexistence of BHs and NSs for the same $M_\mathrm{CO}$. If $M_\mathrm{CO} < M_\mathrm{CO}^\mathrm{crit}$, a NS forms; if not, the compact remnant is a direct BH.\footnote{In context of the BPS code \textsc{sevn} \citep{iorio}, a fallback BH formation window is inserted over $M_\mathrm{CO}$ values in--between the NS and direct BH regimes.}

\citetalias{Mapelli2020} differs from our framework in three principal regards. 
First, the stellar progenitors considered in \citetalias{Mapelli2020} are the single star models from \cite{LC2018} evolved from ZAMS up to the onset of iron-core infall over a parameter space spanned by $M_\mathrm{ZAMS}$, $Z$ and initial rotation $v_\mathrm{ini}$.
Second, as pre-SN explodability criterion, \citetalias{Mapelli2020} uses $\xi_\mathrm{2.5}$. 
Third, the way to relate CCSN outcomes to $M_\mathrm{CO}$ in \citetalias{Mapelli2020} is by coarse-graining the $\xi_\mathrm{2.5}$ values over the sampled $(M_\mathrm{ZAMS}, Z, v_\mathrm{ini})$--parameter space, and then fitting these as a function of $M_\mathrm{CO}$ with a monotonically increasing parametric power-law model. 

The critical compactness value $\xi_\mathrm{2.5}^\mathrm{crit}$ for BH formation is a free parameter in \citetalias{Mapelli2020}. Setting the threshold for BH formation to $\xi_\mathrm{2.5}^\mathrm{crit} = 0.3$, as is suggested in \citetalias{Mapelli2020}, implies $M_\mathrm{CO}^\mathrm{crit} = 4.4 \, M_\odot$. With greater threshold values $\xi_\mathrm{2.5}^\mathrm{crit} \in [0.32, 0.33, 0.37, 0.45]$, the resulting critical CO core masses are $M_\mathrm{CO}^\mathrm{crit}/M_\odot \in [4.8, 5, 6.1, 11]$. In Fig.~\ref{fig:summary}, \citetalias{Mapelli2020} is evaluated for $\xi_\mathrm{2.5}^\mathrm{crit} = 0.33$.

\citetalias{Mapelli2020} and \citetalias{Mandel2020} both are based on stellar evolution tracks that exhibit a lower compactness peak ($\xi_\mathrm{2.5}^\mathrm{max} < 0.45$) at intermediate $M_\mathrm{CO} < 7 \, M_\odot$ (see Fig.~1 in \cite{Mapelli2020} for \citetalias{Mapelli2020}, and Fig.~\ref{fig:heger-schneider-comp} for \citetalias{Mandel2020}), which does not reach the upper threshold for direct BH formation, according to our $\xi_\mathrm{2.5}$--based criterion for BH formation. 

\subsubsection{F12}
In \cite{Fryer2012}, recipes are formulated to compute compact remnant masses. The original recipe uses parametric fits to predict the remnant masses as a function of $M_\mathrm{ZAMS}$ and $Z$ of single stars. It is constructed based on hydrodynamical simulation outcomes and stellar progenitors at two reference metallicities (solar and Population~III) from \cite{Woosley2002}. The original recipe is then reformulated as a function of $M_\mathrm{CO}$  and final pre--SN mass $M_\mathrm{final}$, to account for differences in outcomes due to different assumptions about wind mass loss and binary mass transfer. 

It is assumed that the amount of fallback onto the PNS (of fixed mass $\simeq 1 \, M_\odot$) depends on the timing of the explosion (a ``fast-convection'' explosion, if it happens within 250 ms after core bounce; a ``delayed-convection'' explosion otherwise), since the accretion rate of the infalling material decreases with time and therefore also the total kinetic energy stored in the convective region between the PNS and the base of the shock. The fast-convection (``rapid'') explosion model and the delayed-convection (``delayed'') explosion models do not explicitly distinguish the remnant type. However, the remnant type can be distinguished implicitly in the rapid model, since it predicts a NS-BH mass gap. According to the rapid \citetalias{Fryer2012}, there is no coexistence of NSs and BHs for the same $M_\mathrm{CO}$. For $M_\mathrm{CO} < 6 \, M_\odot$, only NSs form. For $6 \leq M_\mathrm{CO}/M_\odot \leq 7$ and $M_\mathrm{CO} > 11 \, M_\odot$, only direct BHs form. For $7 < M_\mathrm{CO}/M_\odot < 11$, stars are predicted to explode and leave behind fallback BHs, with a fallback mass fraction that increases with $M_\mathrm{CO}$.

\subsubsection{PS20} 
\cite{PS20}  evolve bare CO cores through the late burning phases over a densely sampled grid in the  $(M_\mathrm{CO}, X_C)$ plane at zero age core--carbon burning up to the onset of iron-core infall. To obtain the final fate landscape over this plane, they suggest to apply \citetalias{Ertl2016} as the default explodability criterion. Thus, final fates are ``looked up'' given the starting point in the $(M_\mathrm{CO}$, $X_C)$ grid, which is interpolated over to get predictions at arbitrary values of interest within the grid boundaries $2.5 \leq M_\mathrm{CO}/M_\odot \leq 10$.
With \citetalias{Ertl2016}, successful and failed SNe coexist over the entire range $2.5 \leq M_\mathrm{CO}/M_\odot \leq 10$, provided that $X_C$ is suitably chosen. The gross trend is that at high $X_C$ and low $M_\mathrm{CO}$, explosions dominate, while failed SNe dominate at low $X_C$ and high $M_\mathrm{CO}$ (see the left panel of Fig.~ \ref{fig:ps20-s2123}). For population synthesis purposes, it therefore has been assumed that all stars with $M_\mathrm{CO} < 2.5 \, M_\odot$  succeed and all stars with $M_\mathrm{CO} > 10 \, M_\odot$ fail to explode \citep[e.g.,][]{Patton2022}.

The motivation for evolving bare CO cores is that after the end of CHeB, the evolution of the core and that of the envelope are largely decoupled. The envelope restructures itself on the thermal timescale and has too little time to readjust to the core whose evolution speeds up after core helium burning due to the enhanced neutrino losses -- it only takes a few thousand years from carbon ignition up to iron-core collapse. The argument therefore is that the evolution of the CO core is not affected by binarity and stellar winds after the end of CHeB.

When evaluating our explodability scheme over the \citet{PS20} grid (see Fig.~\ref{fig:ps20-s2123}), we lift the $M_\mathrm{CO}$--based criterion for the following reason. The \cite{PS20} parameter space in the $(M_\mathrm{CO}, X_C)$ plane is sampled homogeneously. With the \citetalias{Schneider21}, \citetalias{Schneider2023}, \citetalias{Schneider2024}, \citetalias{Temaj} and H16 stellar models used for formulating our pre-SN explodability criteria, the $M_\mathrm{CO}$ range is homogeneously sampled only within a sub-range in $X_C$ (see Fig.~\ref{fig:single-stripped}). 
Therefore, to remain agnostic about CCSN outcomes at arbitrary $X_C$, the condition that for $M_\mathrm{CO} < M_\mathrm{CO}^\mathrm{min}$ only explosions occur, is not imposed here. This is tolerable, since the predictions with our pre-SN explodability scheme remain faithful without the $M_\mathrm{CO}$-based criterion, achieving an accuracy of 98.7\% agreement with \citetalias{M16}.

The CCSN recipes are indirectly testable by comparison to observations of compact remnant masses. \citetalias{Mapelli2020} and \citetalias{Mandel2020} both do not predict a BH-BH (``upper'') mass gap, because direct BH formation outcomes are not interrupted after having set in at sufficiently large $M_\mathrm{CO}$. Our CCSN recipe is compatible with a BH-BH mass gap, since there is a SN window for $M_\mathrm{CO}/M_\odot \in (8.4, 12.4)$ independent of MT history and metallicity $Z>Z_\odot/10$, over which the expected outcome is a NS or a fallback BH of lower mass than that of a direct collapse BH at the same $M_\mathrm{CO}$. The default \citetalias{Mapelli2020} model (without fallback) predicts a NS-BH (``lower'') mass gap because of a sharp transition between remnant types at a critical $M_\mathrm{CO}^\mathrm{crit}$ and because low-$M_\mathrm{CO}$  BH progenitors are weakly affected by stellar winds \citep{Mapelli2020}.  \citetalias{Mandel2020} does not predict a lower mass gap, since it predicts  direct BHs, fallback BHs and NSs to coexist over $2 < M_\mathrm{CO}/M_\odot < 7$. 

\subsection{Luminosities of the missing red supergiants}
\label{app:mRSG_L}

The missing RSG problem is typically defined not only in terms of the range in $\log L_\mathrm{pre-SN,obs}$ over which no Type~IIP SN progenitors are observed, but also in terms of the discrepancy between the observed and the expected $\log L_\mathrm{pre-SN,obs}$ distribution, where the expected distribution is weighted by the initial mass function \citep[e.g.,][]{Smartt2009, Smartt2015, DaviesBeasor2018b, Rodriguez2022}.

The conversion of $\log L_\mathrm{pre-SN,obs}$ to $M_\mathrm{ZAMS}$ is strongly dependent on the stellar models, which can yield very different $M_\mathrm{CO}$ for the same $M_\mathrm{ZAMS}$ (see Fig.~\ref{fig:heger-schneider}). In this work, we therefore suggest to define and address the problem in terms of $M_\mathrm{CO}$-- rather than $M_\mathrm{ZAMS}$- ranges.

The luminosity of the brightest, (most likely) core-helium burning stars that have been observed and define the HD limit is lower than that of actual post core-carbon burning SN progenitors. Therefore, $\log L/L_\odot \simeq 5.5$ inferred in \cite{DaviesBeasor2018b} is in fact a lower bound on the upper boundary to the luminosity range over which RSGs are missing.

The lower boundary of the missing RSG luminosities is uncertain. According to \cite{Maund2015}, the inferred $\log L_\mathrm{pre-SN,obs}$ of the SN2009kr progenitor is likely an upper limit, since a compact stellar cluster was interpreted to be contributing to the brightness at the same sky location from the comparison of pre-explosion and late-time photometry. For $\log L_\mathrm{pre-SN,obs}$ of the SN2009hd progenitor, \cite{Elias-Rosa2011} report that its magnitude is close to their empirically derived detection limit, and should therefore rather be considered as an upper limit. In more recent work compiling Type~II SN progenitor luminosity inferences from multiple reconstruction methods, no conclusion is drawn regarding the maximal $\log L_\mathrm{pre-SN,obs}$ \citep{Rodriguez2022}. Therefore, while we adopt the most optimistic assumption about progenitor luminosities following \cite{DaviesBeasor2018b}, we note that the most luminous Type~IIP progenitor is likely fainter than $\log L_\mathrm{pre-SN,obs}^+/L_\odot = 5.36$. 

On the other hand, recent work suggests that the SN progenitor luminosities are systematically underestimated by common assumptions made in the progenitor studies \citep{Beasor2025}. 

\subsection{S21 versus H16 single star models}
\label{app:h16_ff}

We further study the differences in the final fate outcomes for single stars due to differences in the adopted stellar evolution physics. To this end, we exemplarily compare the H16 and the \citetalias{Schneider21} single star models at $Z=Z_\odot$.

\begin{figure}
      \centering
      \includegraphics[width=0.49\textwidth]{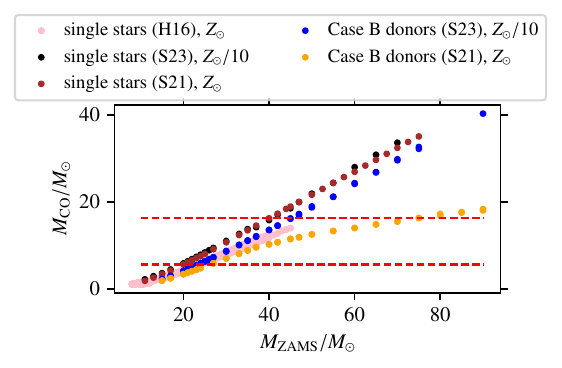}
      \caption{Comparison of the $M_\mathrm{ZAMS}$--to--$M_\mathrm{CO}$ relations of the \citetalias{Schneider21} and the H16 single star models at $Z=Z_\odot$. For the same ZAMS mass, the H16 models result in a substantially lower $M_\mathrm{CO}$ compared to \citetalias{Schneider21}. The red horizontal lines indicate the $M_\mathrm{CO}^\mathrm{min}$ and $M_\mathrm{CO}^\mathrm{max}$ boundaries. Also shown are the $M_\mathrm{ZAMS}$--to--$M_\mathrm{CO}$ relations for single stars at $Z=Z_\odot/10$ (\citetalias{Schneider2023}) and those of Case~B donors at $Z=Z_\odot$ (\citetalias{Schneider21}) and $Z=Z_\odot/10$ (\citetalias{Schneider2023}).} 
    \label{fig:heger-schneider}
\end{figure}

Fig.~\ref{fig:heger-schneider} shows the $M_\mathrm{ZAMS}$-to-$M_\mathrm{CO}$ relations for several suites of single and binary-stripped star models considered in this work. The H16 and the \citetalias{Schneider21} single stars, despite starting from the same $M_\mathrm{ZAMS}$ and the same $Z$, yield remarkably different $M_\mathrm{CO}$ values at the end of CHeB.

\begin{figure}
      \centering
      \includegraphics[width=0.47\textwidth]{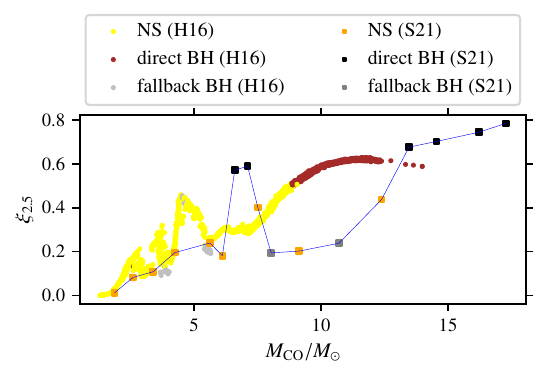}
      \caption{Compact remnants left behind the $Z_\odot$ single star SN progenitor models \citetalias{Schneider21} and H16, predicted by the \citetalias{M16} SN model, are compared with the trends of progenitor $\xi_\mathrm{2.5}$ with $M_\mathrm{CO}$. 
      The $\xi_{2.5}$--peaks are shifted towards higher $M_\mathrm{CO}$ in \citetalias{Schneider21} compared to H16. The peaks are at $\simeq 4.8 \, M_\odot$ in H16 but at $\simeq 7 \, M_\odot$ in \citetalias{Schneider21}, and at $\simeq 9 \, M_\odot$ in H16 but at $\simeq 13 \, M_\odot$ in \citetalias{Schneider21}.
      For improved visual discrimination, the sparsely sampled \citetalias{Schneider21} models are connected by lines (in blue). 
      }     
    \label{fig:heger-schneider-comp}
\end{figure}

H16 comprises a densely sampled grid of models over $2 < M_\mathrm{CO}/M_\odot < 15$. It, therefore, is ideal to study phenomenologically the distribution of remnant types (NS, direct BH, fallback BH) as a function of $M_\mathrm{CO}$. Two important conclusions can be drawn from it (see Fig.~\ref{fig:heger-schneider-comp}): 
\begin{enumerate}
    \item Fallback BH formation is not randomly distributed over the $M_\mathrm{CO}$ range but is clustered, leaving regions of NS formation in-between the clusters.
    \item Fallback BH remnants coexist with NS remnants over clustered $M_\mathrm{CO}$ ranges. 
\end{enumerate}

In H16, in contrast to the \citetalias{Schneider21} single star models, there is no window at intermediate $M_\mathrm{CO}$ (i.e., around the first peak in $\xi_\mathrm{2.5}$) over which BHs would be predicted to form (see Fig.~(\ref{fig:heger-schneider-comp}). The first peak in $\xi_\mathrm{2.5}$ is not large enough to surpass $\xi_\mathrm{2.5}^\mathrm{max}$ for having direct BH formation guaranteed. With the exception of one model at $M_\mathrm{CO} \simeq 2.8 \, M_\odot$, all SN progenitors up to $M_\mathrm{CO} \simeq 9 \, M_\odot$ explode, and BHs form by direct collapse for values beyond. The plateau of direct collapse BH outcomes sets in at a much greater value of $M_\mathrm{CO}^{(3)} = 13 \, M_\odot$ in \citetalias{Schneider21}. In H16, $\xi_\mathrm{2.5}$ peaks at $M_\mathrm{CO} \simeq 4.8 \, M_\odot$, with a $\xi_\mathrm{2.5}$--value substantially lower and the peak position in $M_\mathrm{CO}$ shifted to{ward a lower value by roughly $2 \, M_\odot$ compared to \citetalias{Schneider21}.  Few fallback BHs form near local minima in $\xi_\mathrm{2.5}$. 

\begin{figure}
      \centering
      \includegraphics[width=0.47\textwidth]{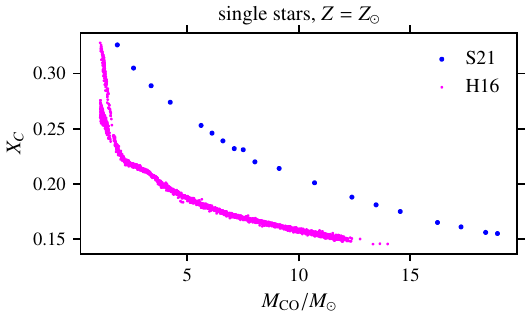}
      \caption{Comparison of the $X_C$--to--$M_\mathrm{CO}$ relations of the H16 and the \citetalias{Schneider21} single star models at the end of CHeB.} 
    \label{fig:heger-schneider-comp2}
\end{figure}

The shift of the compactness peaks toward lower $M_\mathrm{CO}$ values in H16 compared to \citetalias{Schneider21} (see Fig. \ref{fig:heger-schneider-comp}) is explained by a lower $X_C$ abundance in the H16 models compared to the \citetalias{Schneider21} models over the same $M_\mathrm{CO}$ mass range (see Fig.~\ref{fig:heger-schneider-comp2}.)}

Given the large differences in the final fate outcomes as a function of $M_\mathrm{CO}$ despite the same starting conditions $M_\mathrm{ZAMS}$ and $Z$, i.e., coming from the adopted stellar evolution model choice, the question arises which single star model is to be preferred. The hypothesis that failed SNe are part of the solution to the missing RSG problem favors the \citetalias{Schneider21} over the H16 stellar models (see Sect.~\ref{sec:comp_obs}).  
H16 does not address the missing RSG problem, since direct collapse BHs do not form for $M_\mathrm{CO} < 9 \, M_\odot$ when using \citetalias{M16} with parameter choice from \cite{Schneider21} as CCSN model.
The $\xi_\mathrm{2.5}$ peak of the H16 stellar models occurs over a progenitor $M_\mathrm{CO} \simeq 4.8 \, M_\odot$, which is contained inside the range, over which SNe have been observed (see Sect.~\ref{sec:comp_obs}). The $\xi_\mathrm{2.5}$--peak coincides statistically with failed SN outcomes for many parameter choices of \citetalias{M16}. 

\end{appendix}
\end{document}